\documentclass[fleqn,usenatbib]{mnras}

\usepackage{newtxtext,newtxmath}
\usepackage{enumitem}
\usepackage[T1]{fontenc}
\usepackage[utf8]{inputenc}

\DeclareRobustCommand{\VAN}[3]{#2}
\let\VANthebibliography\thebibliography
\def\thebibliography{\DeclareRobustCommand{\VAN}[3]{##3}\VANthebibliography}

\usepackage{graphicx}	% Including figure files
\usepackage{amsmath}	% Advanced maths commands
\usepackage{caption}
\usepackage{subcaption}
\usepackage{multirow}
\usepackage{booktabs}

\title[GOTO RNN classifier]{Light curve classification with recurrent neural networks for GOTO: dealing with imbalanced data}

\author[U. F. Burhanudin et al.]{
U. F. Burhanudin,$^{1}$\thanks{E-mail: ufburhanudin1@sheffield.ac.uk}
J. R. Maund,$^{1}$
T. Killestein,$^{2}$
K. Ackley,$^{3,4}$
M. J. Dyer,$^{1}$
J. Lyman,$^{2}$
K. Ulaczyk,$^{2}$
\newauthor
R. Cutter,$^{2}$
Y.-L. Mong,$^{3,4}$
D. Steeghs,$^{2,4}$
D. K. Galloway,$^{3,4}$
V. Dhillon,$^{1, 11}$
P. O'Brien,$^{5}$
G. Ramsay,$^{6}$
\newauthor
K. Noysena,$^{7}$
R. Kotak,$^{8}$
R. P. Breton,$^{9}$
L. Nuttall,$^{10}$
E. Pall\'e,$^{11}$
D. Pollacco,$^{2}$
E. Thrane,$^{3}$
S. Awiphan,$^{7}$
\newauthor
P. Chote,$^{2}$
A. Chrimes,$^{2}$
E. Daw,$^{1}$
C. Duffy,$^{6}$
R. Eyles-Ferris,$^{5}$
B. Gompertz,$^{2}$
T. Heikkil\"a,$^{8}$
P. Irawati,$^{7}$
\newauthor
M. R. Kennedy,$^{9}$
A. Levan,$^{2}$
S. Littlefair,$^{1}$
L. Makrygianni,$^{1}$
D. Mata-S\'anchez,$^{9}$,
S. Mattila,$^{8}$
\newauthor
J. McCormac,$^{2}$
D. Mkrtichian,$^{7}$
J. Mullaney,$^{1}$
U. Sawangwit,$^{7}$
E. Stanway,$^{2}$
R. Starling,$^{5}$
P. Str\o{}m,$^{2}$
\newauthor
S. Tooke,$^{5}$
K. Wiersema$^{2}$
\\
\\
$^{1}$Department of Physics and Astronomy, University of Sheffield, Sheffield S3 7RH, UK\\
$^{2}$Department of Physics, University of Warwick, Gibbet Hill Road, Coventry CV4 7AL, UK\\
$^{3}$School of Physics \& Astronomy, Monash University, Clayton VIC 3800, Australia\\
$^{4}$OzGRav-Monash, School of Physics and Astronomy, Monash University, Victoria 3800, Australia\\
$^{5}$School of Physics \& Astronomy, University of Leicester, University Road, Leicester LE1 7RH, UK\\
$^{6}$Armagh Observatory \& Planetarium, College Hill, Armagh, BT61 9DG\\
$^{7}$National Astronomical Research Institute of Thailand, 260 Moo 4, T. Donkaew, A. Maerim, Chiangmai, 50180 Thailand\\
$^{8}$Department of Physics \& Astronomy, University of Turku, Vesilinnantie 5, Turku, FI-20014, Finland\\
$^{9}$Jodrell Bank Centre for Astrophysics, Department of Physics and Astronomy, The University of Manchester, Manchester M13 9PL, UK\\
$^{10}$University of Portsmouth, Portsmouth, PO1 3FX, UK\\
$^{11}$Instituto de Astrof'{i}sica de Canarias, E-38205 La Laguna, Tenerife, Spain
}

\date{Accepted XXX. Received YYY; in original form ZZZ}

\pubyear{2020}

\begin{document}
\label{firstpage}
\pagerange{\pageref{firstpage}--\pageref{lastpage}}
\maketitle

\begin{abstract}

The advent of wide-field sky surveys has led to the growth of transient and variable source discoveries. The data deluge produced by these surveys has necessitated the use of machine learning (ML) and deep learning (DL) algorithms to sift through the vast incoming data stream. A problem that arises in real-world applications of learning algorithms for classification is imbalanced data, where a class of objects within the data is underrepresented, leading to a bias for over-represented classes in the ML and DL classifiers. We present a recurrent neural network (RNN) classifier that takes in photometric time-series data and additional contextual information (such as distance to nearby galaxies and on-sky position) to produce real-time classification of objects observed by the Gravitational-wave Optical Transient Observer (GOTO), and use an algorithm-level approach for handling imbalance with a focal loss function. The classifier is able to achieve an Area Under the Curve (AUC) score of 0.972 when using all available photometric observations to classify variable stars, supernovae, and active galactic nuclei. The RNN architecture allows us to classify incomplete light curves, and measure how performance improves as more observations are included. We also investigate the role that contextual information plays in producing reliable object classification.

\end{abstract}

\begin{keywords}
methods: data analysis -- techniques: photometric -- survey
\end{keywords}

%%%%%%%%%%%%%%%%%%%%%%%%%%%%%%%%%%%%%%%%%%%%%%%%%%

%%%%%%%%%%%%%%%%% BODY OF PAPER %%%%%%%%%%%%%%%%%%

\section{Introduction}

Time domain astronomy concerns the study of astronomical objects that exhibit variability in brightness over short timescales, ranging from seconds to months, compared to typical cosmic timescales that span millions to billions of years. The study of transient and variable sources allows us to gain a better understanding of the  Universe, by way of measuring cosmic distances and cosmic expansion. The discovery of the period-luminosity relation of Cepheid variables in the Small Magellanic Cloud enabled the use of Cepheid variables as a standard candle to measure extra-galactic distances \citep{1912Leavitt}. Type Ia supernovae have been used to trace the expansion history of the Universe, and two teams independently measured the luminosity-distance relationship of SNe Ia up to redshift $z\sim 1$ to determine that the Universe is undergoing accelerated expansion \citep{1998riess, 1999Perlmutter}.

Historically, the discovery of transient and variable sources has been serendipitous. The advent of repeated sky surveys such as the Catalina Real-Time Survey \citep[CRTS;][]{2009drake}, the Panoramic Survey Telescope and Rapid Response System \citep[Pan-STARRS;][]{2010kaiser}, the Palomar Transient Factory \citep[PTF;][]{2009rau}, the All Sky Automated Survey for SuperNovae \citep[ASAS-SN;][]{2014shappee}, and the Subaru Hyper Suprime-Cam transient survey \citep[HSC;][]{2019yasuda} have greatly increased the rate at which new transient and variable sources are discovered. The large quantity of data produced by these surveys have led to the discovery of new types of transients such as superluminous supernovae \citep{2007Quimby}, and calcium-rich transients \citep{2012kasliwal}.

Recently, the growing number of surveys across the world has facilitated a new era of multi-messenger astronomy. Observing a single event across multiple wavelengths and through different detectors allows for a deeper understanding of the physics behind transient phenomena. In 2017, the gravitational wave signals of a binary neutron star merger (designated GW170817) were detected by the Advanced LIGO and Advanced Virgo gravitational-wave detectors \citep{2017abbotGW}. Follow-up observations across the electromagnetic spectrum led to the discovery of the kilonova AT2017gfo, thought to be powered by the radioactive decay of $r$-process nuclei following a binary neutron star merger \citep{2017abbotMM, 2017chornock_2017gfo, 2017coulter2017gfo, 2017drout2017gfo, 2017shappee2017gfo, 2017smartt, 2017villar_207gfo}. 

Current surveys such as the Zwicky Transient Factory \citep[ZTF;][]{2018zwicky} and the upcoming Vera C. Rubin Observatory Legacy Survey of Space and Time \citep[LSST;][]{2008lsst} will generate large amounts of observational data. On a typical night of observing, ZTF will produce up to $\sim 1$TB of raw image data, and up to 2 million source alerts extracted from difference imaging \citep{2018Masci}. The rate of data generation exceeds human capability for manual processing, and the role of identifying and classifying new sources from survey data falls to machine learning algorithms \citep{2012bloomrichardsBook, 2012bloom, 2010ballbrunner}

The first step in the automated process of discovery from image data is determining whether a source is astrophysically real, or if it is a 'bogus' detection such as an image artifact (e.g. as a result of poor subtraction from difference imaging) or a cosmic ray. The real-bogus classification problem has been well studied and applied to surveys, and is becoming a standard part of automated discovery pipelines \citep{2021killestein, 2020mong, 2019duev, 2018lin, 2017gieske, 2015wright, 2013brink}.

Once a source has been identified as real, the next step is classifying the source. Various taxonomies exist in astronomy, with broad categorisations such as consistently varying sources versus transient events, and classifications based on spectroscopic features for supernovae subtypes \citep{1997filippenko}. Usually, confidently classifying new optical transients relies on additional observations with spectroscopic facilities. However, in the era of large scale sky surveys, spectroscopic follow-up of all new discoveries is not guaranteed and is prohibitively time-consuming. As a result, alternative methods of classification relying on photometric and image data have been developed.

Both traditional feature-based machine learning (ML) and deep learning (DL) methods have been used to classify astronomical objects into distinct classes. Features are metrics derived from the data that are able to encapsulate the properties that differentiate between different classes. Light curve-derived features have been used to classify a set of simulated supernova light curves from the Supernova Photometric Classification Challenge (SPCC) \citep{2010kessler} into different spectroscopic types \citep{2016lochner, 2013ishida}, and also for supernova classification with real survey data \citep{2020dauphin, 2020Hosseinzadeh}.

In contrast to feature-based ML models, DL models are able to learn salient features from the data, and do not require a feature extraction step prior to training. Deep neural network and recurrent neural network (RNN) architectures have been used to classify simulated light curves \citep{2020moller, 2019Apasquet, 2017charnock} and real light curves \citep{2020takahashi} for supernova classification, general transient classification \citep{2019muthukrishna}, and variable star classification with real light curves \citep{2020becker, 2019tsang, 2017mahabal}. Work has also been done on classifying objects using image stamps as input to convolutional neural networks \citep{2020wardega, 2020gomez, 2019carrasco}.

In search of the optical counterparts to gravitational wave signals is the Gravitational-wave Optical Transient Observer (GOTO) (Steeghs et al. in prep) survey. When GOTO receives an alert for gravitational wave or gamma-ray burst detection from other facilities, it will rapidly begin observing the localised region of sky to look for optical counterparts \citep{2020dyer}.

While not in gravitational wave follow-up mode, GOTO conducts an all-sky survey to search for transient and variable sources. A real-bogus classifier \citep{2021killestein} first identifies astrophysically real detections and 'bogus' subtraction artefacts from difference images. Detections that are identified as real by the real-bogus classifier are then passed on to the GOTO Marshall (Lyman et al. in prep) , a web-based interface for GOTO observers to vet, search, and trigger follow-up observations of new discoveries. The next step in object classification is classifying real discoveries into distinct astrophysical types. Providing object classifications for real objects will be useful for the GOTO collaboration in helping to identify interesting targets for follow-up and further science goals.

Effective classification by ML and DL models rely on good representation of the labelled classes in the data set to learn class separability across the labelled objects. In real-world applications, the data will typically contain one or more classes that have more examples than other classes. This type of data is referred to as 'imbalanced data', and it poses a difficulty for classification as models will be biased towards the class where there are many more examples to learn from. Here, we present an RNN-based classifier for classification of objects discovered by GOTO. 

In section \ref{sec:goto_data}, we provide an overview of the GOTO survey and the data used to train and test the RNN classifier. In section \ref{sec:model}, we introduce the RNN architecture, the class imbalance problem, and the approach taken to deal with an imbalance data set. In section \ref{sec:method}, we outline the training process, and in section \ref{sec:results}, we discuss the performance of the classifier, and how contextual information plays a role in how models learn to classify. We discuss how the work presented in this paper can be improved upon, and conclude with sections \ref{sec:discussion} and \ref{sec:conclusion}.

\section{The GOTO survey and data}
\label{sec:goto_data}

\subsection{The Gravitational-wave Optical Transient Observer}
\label{sec:goto}

The Gravitational-wave Optical Transient Observer (GOTO) is a ground-based observatory, with a modular design situated at the Roque de los Muchachos Observatory on La Palma, Canary Islands (Steeghs et al. in prep). GOTO consists of multiple nodes, with each node hosting an array of up to eight 40 cm diameter unit telescopes (UTs) providing a combined 40 square degree field of view in a single pointing. The current configuration consists of a single node in La Palma, with a plan to add another node in La Palma and another two nodes at the Siding Spring Observatory in Australia. When fully complete, GOTO will have two nodes in La Palma (GOTO North) and two nodes in Australia (GOTO South), for a total of 4 nodes and 32 UTs and the ability to have constant coverage of the night sky. The primary science aim of GOTO is to search for optical signatures following a gravitational wave detection from detectors such as the Advanced LIGO \citep{ligo2007} and Advanced Virgo \citep{2015virgo} facilities. When GOTO is not searching for gravitational wave counterparts, it conducts an all-sky survey enabling the discovery of new variables and, in particular, new transients.

Each GOTO UT is equipped with four Baader filters: a wide band $L$ filter (covering $400-700~\mathrm{nm}$), and three $R,G,B$ filters. GOTO currently conducts all-sky surveys with the $L$ filter, and is able to achieve a depth of up to $20.5$ mag in a 60 second exposure under dark conditions. New transient and variable detections are obtained through difference imaging, where the difference between an image of a new source that is undergoing a change in brightness and a reference image at the same location is calculated to produce an image of the new source. The difference imaging process is done through an image processing pipeline (Steeghs et al. in prep).

\subsection{Data}
\label{sec:data}

The data set used to train the classifier is obtained from GOTO observations spanning the period of 20 March 2019 to 4 November 2020, during the GOTO prototype phase. Light curves of objects observed during this period are created using photometric measurements derived from difference imaging in the $L$ filter. The catalog of GOTO objects is then cross-matched to a number of external catalogs to determine objects that have also been observed by other telescopes and surveys, and to obtain classification labels. The list of external catalogs include:

\begin{enumerate}[label=\arabic{enumi}.,ref=\arabic{enumi}, leftmargin=*]
    \item The American Association of Variable Star Observers (AAVSO) International Variable Star Index \citep{2006watson}
    \item The Veron Catalog of Quasars \& AGN, 13th edition \citep{2010veron}
    \item The Transient Name Server \footnote{https://wis-tns.weizmann.ac.il/}
\end{enumerate}

Additionally, the GOTO objects are also cross-matched against the Galaxy List for the Advanced Detector Era (GLADE) \citep{2018dalya} galaxy catalog to identify if there is a nearby galaxy associated with the object. In total, the data set comprises $99,201$ labelled objects, and are split into three broad classes: variable stars (VS), active galactic nuclei (AGN), and supernovae (SN). The data set is heavily imbalanced with $99~\%$ $(98,457)$ of labelled objects belonging to the variable star class, and only 543 and 201 belonging to the active galactic nuclei and supernovae classes, respectively. The largest class (VS) contains almost 500 times more examples than the smallest class (SN).

Within the astronomical taxonomy for transient and variable sources, there are wide range of classification schemes: classifying variable stars by the physical mechanism that causes variability (eclipsing binaries, RR Lyrae stars, Cepheids) and classifying supernovae by spectroscopic features (type Ia, type Ib/c, type II). The use of `super-classes' that group together distinct types of objects simplifies the classification task while still providing clear classifications. For rapidly evolving objects that would benefit from early time follow-up observations such as supernovae, it is beneficial to separate it from objects that show photometric variation over longer timescales with just a few observations. Providing a more general classification in real-time acts as a `first-pass' classification, and further classification into more specific subtypes can be done when additional observations become available.

\begin{figure}
    \centering
    \includegraphics[width=0.45\textwidth]{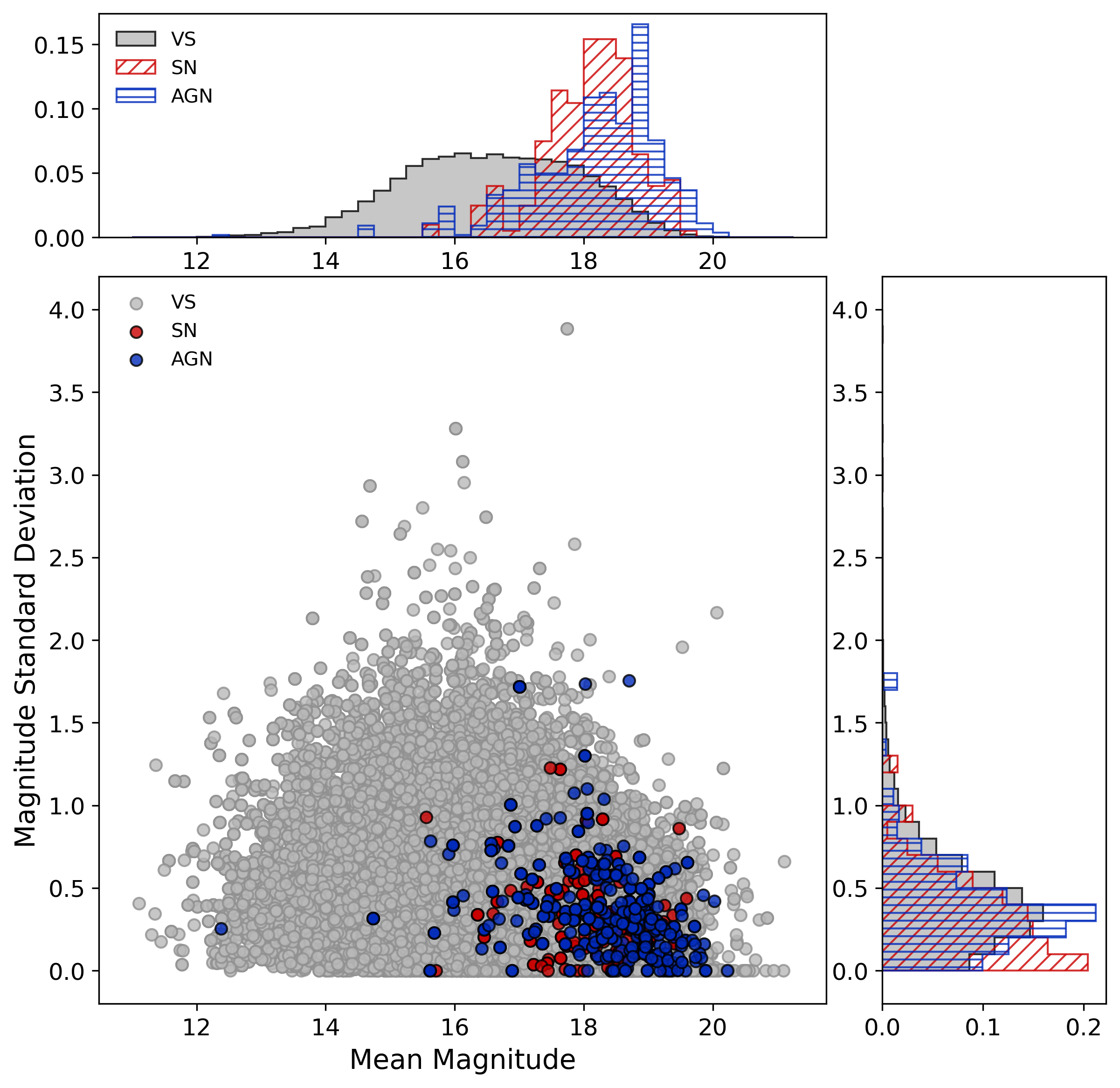}
    \caption{A scatter plot of the mean magnitude of light curves against the standard deviation in magnitudes of light curves for variable stars (VS) in grey, supernovae (SN) in red, and active galactic nuclei (AGN) in blue. In the top and left panels are histograms showing the distribution of mean magnitudes (top) and standard deviation in magnitudes (left). The histograms are plotted as normalised counts, for each class.}
    \label{fig:mag_stdmag}
\end{figure}

\begin{figure}
    \centering
    \includegraphics[width=0.45\textwidth]{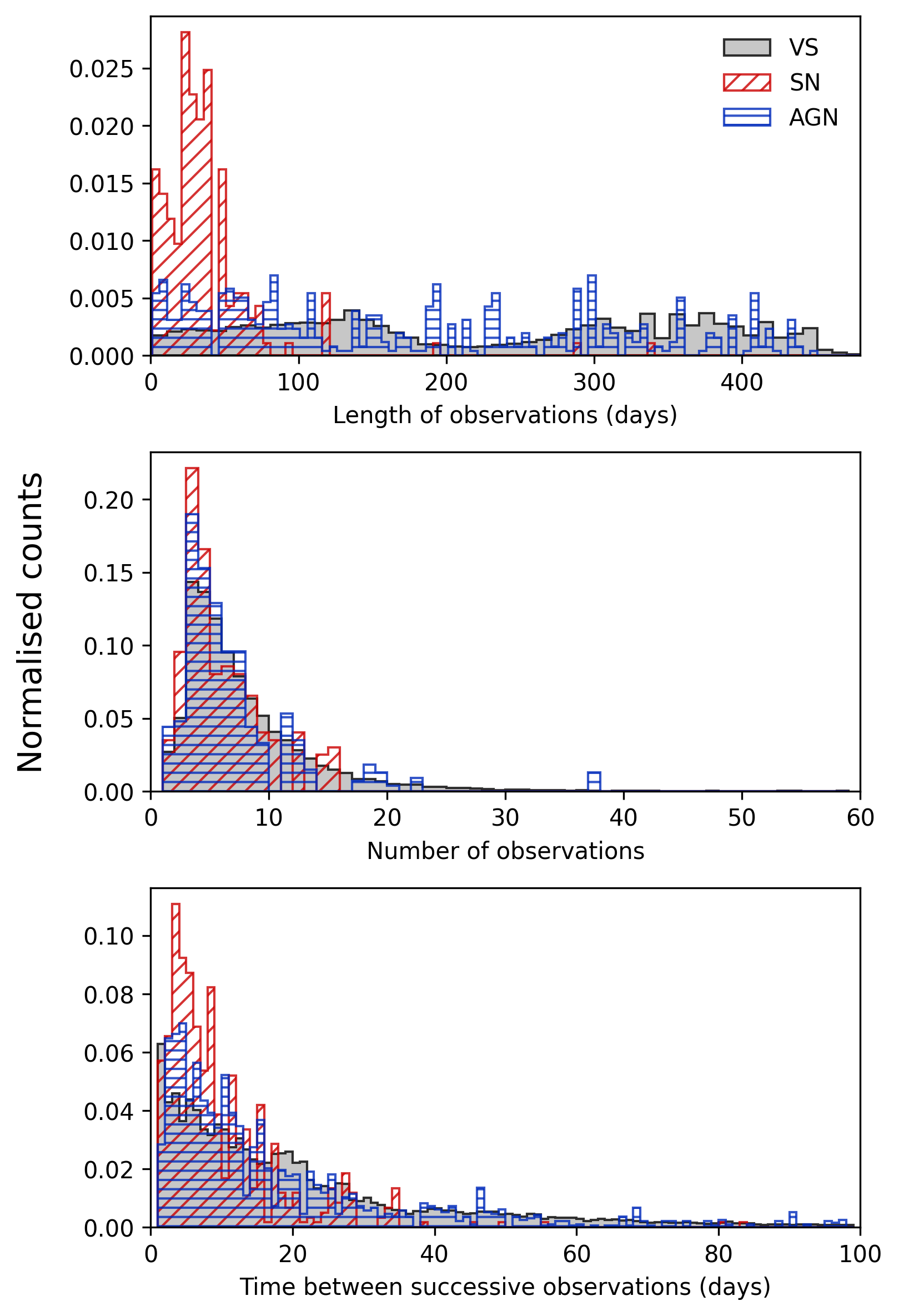}
    \caption{Normalised histograms showing different properties of the light curves in the data set, separated by class. From top to bottom: the time between the first and last observations of the light curves (length of observation), the number of observations in the light curves, and the time between successive observations over all light curves.}
    \label{fig:lc_statistics}
\end{figure}

Figure \ref{fig:mag_stdmag} shows the distribution of mean magnitudes and standard deviation in magnitudes, where each statistic is calculated from all measurements in a light curve. VS have a broad range of magnitudes, while SN and AGN tend to be fainter compared to VS. Figure \ref{fig:lc_statistics} summarises the light curve properties within the data set. SN are typically observed no longer than 100 days, while VS and AGN are observed over longer timescales (>300 days). The median number of points in a light curve for the data set is 6, and the median time between successive observations is $\sim~11$ days. The data set contains light curves ranging in duration from a single day to a few hundred days. The distribution of number of observations in a light curve and the time between successive observations within a light curve over all classes is fairly similar.

\subsection{Data preprocessing}
\label{subsec:preprocessing}

Before being used as input into the classifier, the data needs to be preprocessed. For each light curve, the time-series input matrix consists of the times of observation $\mathbf{t}$, the magnitudes $\mathbf{m}$, and the errors in magnitude $\pmb{\sigma}_m$. The time is scaled to the time of the first observation, so that it starts at zero, and subsequent time steps are times since the first observation. The time-series input matrix $\mathbf{X_T}$ for a light curve with $n$ observations is

\begin{equation}
    \mathbf{X_T} = 
    \begin{bmatrix}
    \mathbf{t} \\
    \mathbf{m} \\
    \pmb{\sigma}_{m}
    \end{bmatrix} =
    \begin{bmatrix}
    t_0 & t_1 & ... & t_n \\
    m_0 & m_1 & ... & m_n \\
    \sigma_{m_0} & \sigma_{m_1} & ... & \sigma_{m_n}
    \end{bmatrix}.
\end{equation}
Light curves with more than 30 observations are truncated, and light curves with fewer than 30 observations are padded, so that all time-series input matrices $\mathbf{X_T}$ are matrices with 5 rows and 30 columns, where each column represents a time step. Padding is required since the input matrices to the classifier needs to have a fixed input size, but a masking layer can be used to tell the classifier to ignore padded time steps (section \ref{subsec:mixed_input}). We choose 30 observations since we are interested in being able to classify objects early in their light curve evolution.

\begin{figure}
    \centering
    \includegraphics[width=0.45\textwidth]{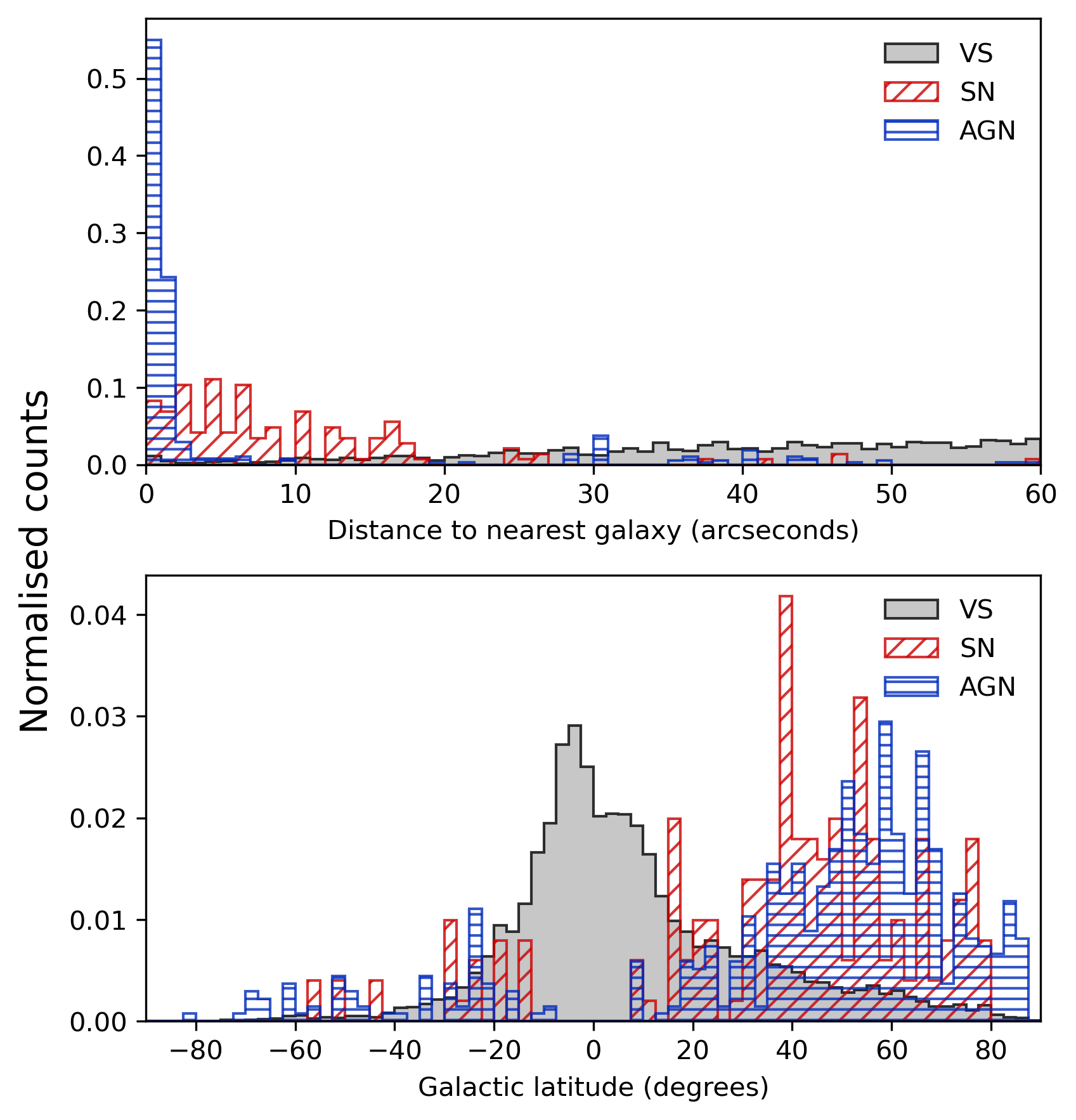}
    \caption{Normalised histograms showing how the distance to the nearest galaxy in the GLADE catalog and galactic latitude for all objects in the data set varies by class. }
    \label{fig:cont_statistics}
\end{figure}

For each object, the contextual information used are the galactic longitude $l$ and latitude $b$ in degrees, and distance in arcseconds to the nearest galaxy in the GLADE catalog $d_G$. Objects that have $d_G > 60~\mathrm{arcseconds}$ have their $d_G$ set to a dummy value. The contextual information input vector $\mathbf{X_C}=(l, b, d_G)$ is used alongside the time-series input matrix $\mathbf{X_T}$ as inputs for the classifier. Figure \ref{fig:cont_statistics} shows the distribution of $d_G$ and $b$ for all objects. AGN and SN are more commonly found to have a nearby galaxy in the GLADE catalog than VS. VS are mostly located close to the galactic plane, while AGN and SN are usually found $\gtrsim~10^{\circ}$ away from the galactic plane.

\section{Model}
\label{sec:model}

\subsection{Recurrent Neural Networks}
\label{subsec:rnn}

Recurrent neural networks (RNNs) are a class of neural networks that operate on sequential data. The sequential data can take the form $[\pmb{x}_{0}, \pmb{x}_{1},\dots, \pmb{x}_{\tau}]$, where $\pmb{x}_{t}$ is a vector with time step index $t$ running from $t=0$ to $t=\tau$ for a sequence with $\tau$ time steps. The time step index $t$ does not necessarily have to represent the passage of time, but can also denote the position of a vector in the sequence. RNNs make use of `hidden states' (similar to the hidden layers of a deep neural network) that incorporate information from the hidden state at the previous time step.

The traditional RNN learns by minimising a loss function (loss functions are explained in detail in section \ref{subsec:imbalance}) through gradient optimisation, in a process known as backpropagation through time \citep{1990werbos} - this is the analog for backpropagation in fixed input size neural networks but applied to sequential data. Traditional RNNs struggle to encapsulate long-term time dependencies because gradients propagated over many time steps tend to explode or vanish - this is known as the \textit{vanishing and exploding gradient problem}. Vanishing gradients make it difficult to improve the cost function because the incremental steps needed to find a local minimum become infinitesimal, and exploding gradients can make training unstable since since a local minimum may never be found. Practical applications in processing sequential data make use of gated recurrent neural networks, which can control the flow of information through time. For an in-depth discussion on recurrent neural networks, refer to \citet{2016goodfellow}.

There are two widely used gated RNN models: the long short-term memory (LSTM) and gated recurrent unit (GRU) networks. LSTM networks \citep{1997hochreiter} use "LSTM" cells that behave similar to hidden states in vanilla RNNs, but have additional parameters and a series of gates that control the flow of inputs and outputs within the cell, and the amount of information from previous timesteps to retain. The cell parameters are trainable, so the LSTM network also learns how to control the flow of information through time. An alternative gated RNN is the GRU \citep{2014cho}, which uses a single gate rather than a set of multiple gates to control how information flows within the unit. 

In this work, both LSTM and GRU networks are used to process light curve data and produce a set of class probabilities. It should be noted that the values returned by the classifier are not automatically true probabilities, rather they are a scores given to an object by the classifier that indicate the level of 'belongingness' to a certain class. In this work, the scores returned by the RNN classifiers are referred to as the class or prediction probabilities. The outputs are produced after reading in the entire light curve, along with additional contextual information.

\begin{figure}
    \centering
    \includegraphics[scale=0.55]{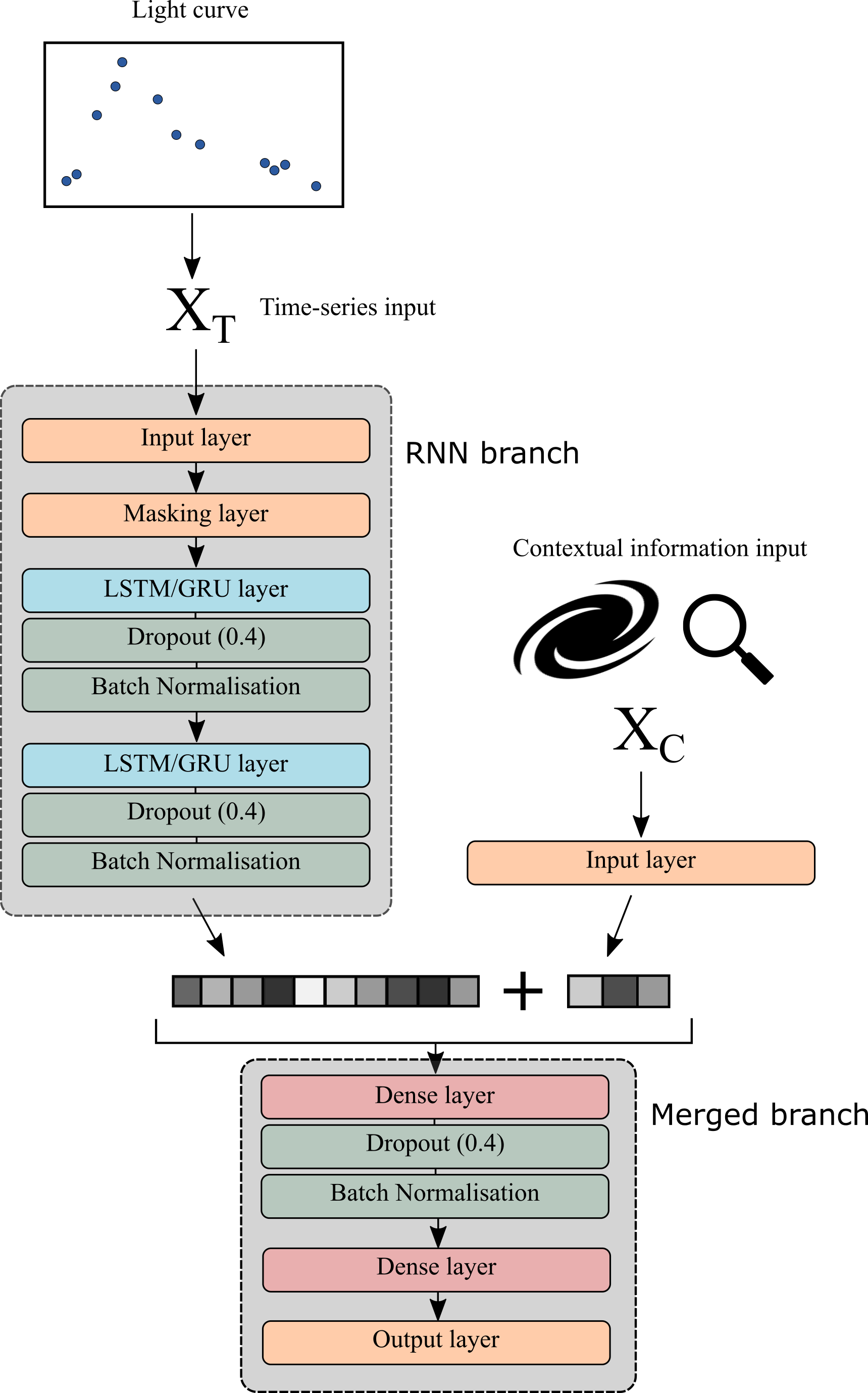}
    \caption{Diagram of the mixed input network. The time-series input $\mathbf{X_T}$ is passed to the RNN branch, and the contextual information input $\mathbf{X_C}$ is appended to the output of the final RNN layer, before being passed on to merged branch of the network.}
    \label{fig:network_diagram}
\end{figure}

\subsection{Mixed input network}
\label{subsec:mixed_input}

In order to utilise both the time-series data from the light curve and contextual information, the neural network model is formed of two branches: the first branch reads in the time-series input matrix $\mathbf{X_T}$ and the second branch reads in the contextual information input vector $\mathbf{X_C}$. Since recurrent neural networks are optimised for processing sequential data, the contextual information is fed to the model separately as opposed to together with the time-series data (e.g. as a vector of constant values), so that the RNN can extract high-level time-series features from the light curve. We use a similar approach to \cite{2019muthukrishna} to construct the RNN branch of the network.

The first branch acts as a standard RNN, consisting of two RNN layers, and the second branch is just an input layer for $\mathbf{X_C}$. The output of the final layer in RNN branch is concatenated with $\mathbf{X_C}$, and forms the input for two dense layers, the latter of which is connected to the final output layer. The output layer produces a list of numbers which sum to 1, which are the class probabilities of an object belonging to each of the defined classes. Figure \ref{fig:network_diagram} illustrates the model architecture. To avoid overfitting (where a model performs well on the training set but underperforms on the test set) dropout and regularisation are used within the model. A summary of the different components of the model is presented below.

\begin{itemize}
\item \textbf{Masking layer}: The masking layer is placed between the input layer of the time-series branch of the model and the first RNN layer. It applies a mask to a sequence of of time steps, where each time step refers to a column in the $\mathbf{X_T}$, and uses a mask value to skip time steps. Subsequent layers will ignore masked timesteps. A masking layer allows the model to process sequences with a different number of time steps.\\

\item \textbf{Long short-term memory (LSTM) layer}: The LSTM layer takes in the masked $\mathbf{X_T}$ as input, and applies the LSTM operation. The dimension of the output $n_{\mathrm{out}}$ from this layer is an adjustable hyperparameter. Two LSTM layers are stacked in the RNN branch of the model: the first LSTM layer returns a sequence of outputs each with dimension $n_{\mathrm{out}}$, which is passed on to the second LSTM layer that returns a single output with dimension $n_{\mathrm{out}}$. \\

\item \textbf{Gated recurrent unit (GRU) layer}: The GRU layer functions in the same way as the LSTM layer and has the same adjustable hyperparameter $n_\mathrm{out}$, but it applies the GRU operation to the data instead. Like the LSTM layers, two GRU layers are stacked and the first GRU layer passes a sequence to the second GRU layer, which returns a single output with dimension $n_\mathrm{out}$.\\ 

\item \textbf{Dropout}: A dropout layer randomly drops input neurons and their corresponding connections during training, as a method to reduce overfitting \citep{srivastava2014}. This forces the neurons to derive more meaningful features from the data without heavily relying on other neurons in the network. During testing, the data is passed through the network without dropout. The dropout fraction sets the fraction of input neurons that are dropped. \\

\item \textbf{Batch normalisation}: During training, the parameters for each layer change, affecting the distribution of the inputs in the proceeding layers in the network. The change in the distribution of inputs in the layers requires the layers to adapt to the new distribution, a phenomenon known as internal covariate shift. Batch normalisation scales the inputs to a layer for each batch so that mean value is close to 0 and the standard deviation is close to 1, reducing the impact of internal covariate shift \citep{2015ioffe}.\\

\item \textbf{Dense layer}: A dense layer is the simplest layer in a neural network: it consists of a fully connected layer of neurons and takes in a fixed size input. The number of neurons in the dense layer is an adjustable hyperparameter. All dense layers in the merged branch have the same number of neurons, which is set by the hyperparameter. The output layer is just a dense layer with three neurons with a softmax activation function that ensures the values returned by the output layer all sum to 1.\\

\item \textbf{Regularisation:} Regularisation introduces a penalty term to the loss function as a method to reduce overfitting. The L2 regularisation is used, which adds a penalty term equal to the sum of all the model weights squared, multiplied by a regularisation factor $\lambda$. The regularisation factor $\lambda$ sets the strength of 'weight decay' in the loss function. A larger value of $\lambda$ forces the weights to have smaller values and helps to reduce overfitting \citep{2016goodfellow}.\\
    
The LSTM and GRU layers serve the purpose of extracting meaningful features from the time-series data. The dense layers then combine the the time-series representations and contextual information and further extract features from the combined features to produce a prediction. After each LSTM/GRU layer and the first dense layer, dropout is applied in an attempt to reduce overfitting, and batch normalisation is applied to reduce internal covariate shift throughout the network. We use a combination of three methods as an approach to deal with potential overfitting: dropout, batch normalisation, and regularisation in the loss function. It is possible to further investigate how these methods work with or without each other and the impact they have on reducing overfitting, but it is beyond the scope of this work.
    
\end{itemize}

\subsection{Class imbalance}
\label{subsec:imbalance}

In classification problems, class imbalance occurs when one class contains significantly fewer examples compared to other classes. The class with fewer examples is often referred to as the minority class, and the classes with many examples are referred to as the majority class. Extreme imbalance can occur when the minority class contains significantly fewer examples than the majority class. Learning from imbalanced data can be difficult, since conventional ML and DL algorithms assume an even distribution of classes within the data set. Classifiers will tend to misclassify examples from the minority class, and will be optimised to perform well on classifying examples from the majority class.

There are three main approaches for dealing with class imbalance \citep{2016Krawczyk}:

\begin{enumerate}[label=\arabic{enumi}.,ref=\arabic{enumi}, leftmargin=*]
\item \textit{Data-level methods}: reduce imbalance by modifying the data with resampling methods
\item \textit{Algorithm-level methods}: modify the algorithm to reduce bias towards examples from the majority class
\item \textit{Hybrid methods}: combine both data resampling and algorithm-level methods
\end{enumerate}

The problem of class imbalance within astronomy has been addressed in a number of works dealing with light curve classification. Synthetic Minority Over-sampling Technique \citep[SMOTE;][]{smote_ref} and a Gaussian resampling variant have been used to augment a training set of spectroscopically classified supernovae from the Pan-STARRS1 Medium-Deep Survey \citep{2019Villar}. Gaussian processes have been used to augment a set of simulated LSST lightcurves from the PLAsTiCC data set \citep{2018PLAsTiCC} to generate a more representative training set \citep{2019Boone}. \cite{2020Hosenie} combine data resampling methods and a hierarchical classification system to deal with class imbalance for variable star classification. 

All the above methods use feature-based ML algorithms to classify light curves. Despite the popularity of DL-based classifiers, there is a lack of research into dealing with class imbalance when using DL architectures \citep{2019Johnson}. Data augmentation for light curves of different astrophysical objects can be a laborious process, as models are needed to generate simulated examples of real observations, and multiple models may be required to simulate objects from multiple classes. This work attempts an algorithm-level approach for dealing with class imbalance by using a focal loss function to optimise the RNN classifier, as an alternative to a data-level approach such as data augmentation. 

In supervised learning, a model is trained for a prediction task by optimising an objective function, commonly referred to as the loss function. A loss function measures the error between the model output and the target output. The model adjusts its internal weights to reduce the error by calculating the gradient in weight space that will minimise the loss function value \citep{2015lecun}. This process is referred to as gradient descent.

For multi-class classification, the cross entropy loss is typically used. Given a multi-class problem with $N$ classes, the cross entropy loss (CE) for an example $i$ is

\begin{equation}
\label{eq:crossentropy}
    \mathrm{CE} = 
    -\displaystyle\sum_{j=1}^{N}\delta_{ij}\log(p_{ij}),
\end{equation}
where $p_{ij}$ is the probability of example $i$ belonging to class $j$, and $\delta_{ij}$ is the Kronecker delta function. The loss for the entire data set is given by summing the loss of all examples.

The focal loss \citep{2017lin} addresses class imbalance by down-weighting examples from the majority class. For the same multi-class problem as above, the focal loss (FL) for an example $i$ is

\begin{equation}
\label{eq:focalloss}
    \mathrm{FL} = 
    -\displaystyle\sum_{j=1}^{N}\delta_{ij}(1-p_{ij})^{\gamma}
    \log(p_{ij}),
\end{equation}
where $(1-p_{ij})$ is the modulating factor, and $\gamma$ is the parameter that adjusts the rate at which majority class examples are down-weighted. Increasing $\gamma$ reduces the contribution from well classified examples to the loss, and increases the importance of improving misclassified examples \citep{2017lin}. For a misclassified example, $p_{ij}$ is small so the modulating factor is close to 1 and the loss is unaffected. Examples that are well classified will have $p_{ij}$ close to 1, so the modulating factor is small, and the loss from well classified examples will be down-weighted.
The focal loss is equivalent to the cross entropy loss when $\gamma=0$. In practice, a weighted version of the focal loss can be used, which \cite{2017lin} find to perform better than the unweighted focal loss (eq. \ref{eq:focalloss}) for imbalanced classification tasks:

\begin{equation}
\label{eq:focalloss_balanced}
    \mathrm{FL} =
    -\displaystyle\sum_{j=1}^{N}\delta_{ij}\alpha_{j}(1-p_{ij})^{\gamma}
    \log(p_{ij}),
\end{equation}
where $\alpha_{j}$ is a weighting factor for class $j$. In this work, the weighting factor is given by

\begin{equation}
    \alpha_{j} = \frac{1}{n}\times\frac{N}{N_{j}}
\end{equation}
where $N$ is the total number of examples in the training set, $N_j$ is the number of examples in class $j$ in the training set, and $n$ is the total number of classes which in this case is $n=3$. For all the above loss functions, the best case is a loss value of zero.

\section{Method}
\label{sec:method}

The GOTO data set is split into 70\% for training, and 30\% for testing. Of the training set, 30\% is set aside for validation. The data are split so that the training, validation, and test sets all have the same proportions of VS, AGN, and SN. A validation set is useful as it provides a measure of how well a model is able to generalise during training. The RNN is implemented and trained using the TensorFlow 2.0 package for Python \citep{2016tensorflow}\footnote{https://www.tensorflow.org/} with Keras \citep{chollet2015keras} for implementation of network layers.

\subsection{Classification metrics}
\label{subsec:metrics}

For a binary classification task where one class is positive and the other class is negative, there are four possible outcomes when evaluating the performance of a classifier. If an example from the positive class is classified as positive, then the result is a \textit{true positive}; if it is classified as negative then the result is a \textit{false negative}. If an example from the negative class is classified as negative, the result is a \textit{true negative}; if it is classified as positive then the result is a \textit{false positive}. 

\begin{figure}
    \centering
    \includegraphics[scale=0.5]{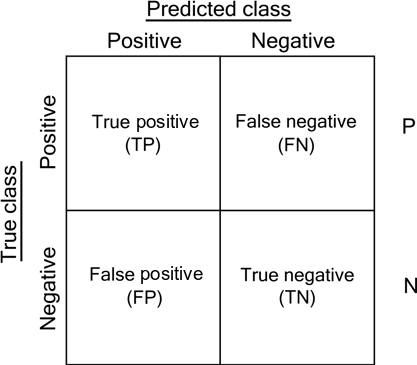}
    \caption{A confusion matrix representing the possible outcomes in binary classification for a positive and negative class. Class labels along the horizontal axis are the labels predicted by the classifier, and class labels along the vertical axis are the true labels. The total number of all true positives is denoted by $P$ and the total number of all true negatives is denoted by $N$.}
    \label{fig:confusion_matrix}
\end{figure}

Figure \ref{fig:confusion_matrix} shows a confusion matrix that represents the possible outcomes from a classifier in a binary classification task. There a number of commonly used classification metrics that can be calculated from the confusion matrix. The accuracy of a classifier is 

\begin{equation}
    \mathrm{Accuracy} = \frac{TP + TN}{P + N}.
\end{equation}
The $F_1$ score of a classifier is

\begin{equation}
    F_1 = \frac{TP}{TP + \frac{1}{2}(FP + FN)}.
\end{equation}
The true positive rate (TPR) or the recall, is

\begin{equation}
    \mathrm{TPR} = \frac{\mathrm{Correctly~classified~positives}}{\mathrm{Total~ positives}} = \frac{TP}{TP + FN} = \frac{TP}{P}.
\end{equation}
The false positive rate (FPR) or the false alarm rate is

\begin{equation}
    \mathrm{FPR} = \frac{\mathrm{Incorrectly~ classified~ negatives}}{\mathrm{Total~ negatives}} = \frac{FP}{FP + TN} = \frac{FP}{N}.
\end{equation}
When dealing with imbalanced data, metrics such as accuracy and $F_1$ score are sensitive to the class distribution in the data set. For example, a classifier predicting on a test set that contains $99\%$ positive examples and $1\%$ negative examples can achieve an accuracy of $99\%$ by predicting all examples as positive. Consider the confusion matrix in figure \ref{fig:confusion_matrix}. The proportion of positive to negative examples is a relationship between the top row (positive) and the  bottom row (negative). Metrics that are calculated using values from both rows (i.e. accuracy and $F_1$ score) will be sensitive to class imbalance. Alternative metrics for classification are receiver operating characteristics (ROC) graphs and the area under the ROC graph (AUC). The ROC is based on the TPR and FPR, where each metric is a ratio calculated from values along a row of the confusion matrix, and hence is insensitive to class imbalance. ROC graphs plot the TPR on the $y$-axis and the FPR on the $x$-axis.

A ROC graph shows the trade-off between true positives and false positives - an ideal classifier will return all true positives and no false positives. A classifier that returns discrete classifications (i.e. one that predicts the class labels) will produce a single pair of values for TPR and FPR, and a corresponding point in ROC space. A classifier that returns probabilities can produce a range of TPR and FPR values by varying a threshold: if the probability is above the threshold, then the classifier returns a positive classification, if it is below the threshold then the classifier returns a negative classification. Since different threshold values produce different TPR and FPR values, it is possible to plot a curve in ROC space by varying the threshold.

In order to compare different classifiers, it is convenient to have a single score that represents classifier performance. Since the ROC curve is in two-dimensional space, the AUC is a fraction of the total area of ROC space, and will always have a value between 0 and 1. For an in-depth discussion of ROC analysis, see \cite{2006fawcett}.

These metrics can extended to a multi-class classification problem: the confusion matrix for $n$ classes becomes an $n\times n$ matrix with the diagonal entries representing correct classifications and the off-diagonal entries representing incorrect classifications. A method for plotting ROC graphs for multiple classes is to plot ROC graphs for each class, treating one class as the positive, and all others as negative. A formulation of the AUC for multi-class classification that is insensitive to imbalanced data was derived by \cite{2001handtill}, which calculates the unweighted mean of the pairwise AUC over all classes. Although this formulation of the AUC is insensitive to class imbalance, it is not straightforward to visualise the ROC space with this method. Nevertheless, the pairwise AUC metric is useful to evaluate the performance of multiple classification models during hyperparameter optimization.

\subsection{Hyperparameters}
\label{subsec:hyper-params}
The model has a number of adjustable hyperparameters, parameters that are not derived through training, but are pre-defined before the training process. Table \ref{tab:hyperparams} lists the adjustable hyperparameters of the model. Two hyperparameters are varied during training for all models: the dimension of the LSTM/GRU output and the number of neurons in the dense layers. We select the values of the LSTM/GRU output dimension to be similar to those used in \citep{2019muthukrishna}. For models trained with the focal loss, there is another hyperparameter that is varied: the focal loss $\gamma$ parameter. We limit the number of adjustable hyperparameters since we want to limit the total number of models to be trained - there are a total of 32 models with all possible hyperparameter combinations. We find that varying the hyperparameters during hyperparameter optimisation does not hugely impact performance on the validation set, and the hyperparameters listed in Table \ref{tab:hyperparams} produce good classification results.

\begin{table}
    \centering
    \begin{tabular}{l|c}
        \toprule
         \textbf{Hyperparameter} & \textbf{Value} \\
         \hline
         Batch size & 128 \\
         Learning rate & $1\times10^{-4}$ \\
         LSTM/GRU output dimension &  $[100, 150]$ \\
         Dropout & 0.4 \\
         Number of neurons in dense layers & $[100, 200]$ \\
         Focal loss $\gamma$ & $[1.0, 2.0]$ \\
         Regularization factor $\lambda$ & 0.01 \\
         \bottomrule
    \end{tabular}
    \caption{Adjustable hyperparameters in the model. Hyperparameters with values in square brackets indicate the range of values used during training.}
    \label{tab:hyperparams}
\end{table}

Hyperparameters that remain fixed are the batch size, the learning rate, the dropout fraction, and the regularization factor. Batch size sets the number of samples that is passed through the model at each epoch of training before the weights are updated. Given that the data is imbalanced, the batch size is set to 128 to ensure that examples from the minority class are passed through the model while the weights are being updated. We note that this is a rudimentary method, and a more thorough approach would be to employ a 'stratified' batching, where each batch has the same proportion of minority and majority examples. Nevertheless, we find that simply setting the batch size to 128 is sufficient to achieve good results (section \ref{sec:results}). The learning rate defines the step size taken during gradient descent to determine the optimal set of weights, in other words it controls how much the weights are changed during training. A learning rate that is too small will fail to find a local minimum in weight space, and a learning rate too large will result in unstable training and fail to find a local minimum. During training, the learning rate is left to the default TensorFlow value of $1\times10^{-4}$. 

\cite{srivastava2014} find that setting the probability of retaining a unit in the network between 0.4 and 0.8 optimally reduces test error on a classification task with a deep neural network. In this work, the dropout fraction used in the TensorFlow implementation is defined as the fraction of units that are dropped, so the optimal range found by \citep{srivastava2014} translates to $0.2 - 0.6$ when expressed as the fraction of units to be dropped. We opt for a dropout fraction of 0.4 in this work. The regularization factor $\lambda$ is set to the default TensorFlow value 0.01. 

The dropout fraction and regularization factor $\lambda$ could have been used as additional hyperparameters to see if varying their values would impact classification performance, but we choose to keep these values constant as to minimise the total number possible model configurations that need to be trained. The main motivation for this work is to see how using weighted loss functions affects performance on imbalanced data.

\subsection{Training process}
\label{subsec:training}

There are two different RNN architectures used, one with LSTM and the other with GRU, and three different loss functions: an unweighted cross entropy loss (eq. \ref{eq:crossentropy}), a weighted cross entropy loss, and a weighted focal loss (eq. \ref{eq:focalloss_balanced}). In total there are six classes of models, each trained with a range of hyperparameters (Table \ref{tab:hyperparams}). For models trained with the cross entropy loss functions, the focal loss parameter $\gamma$ is not an adjustable parameter.

All models are trained for 200 epochs with no early stopping using the Adam optimizer \citep{kingma2014}, and then evaluated on the validation set. The best performing model is selected by choosing the model that has the best AUC score calculated on the validation set. The best models from the six different configurations are then evaluated on the test set. Training was executed on an 8-core CPU. The average time taken to train a single model was approximately one hour at an average of 20 seconds per epoch, and the total time taken to train all models presented in this paper was $\sim44$ hours. The trained models are able to return predictions on the entire test set within $\sim5~\mathrm{seconds}$.

\begin{table*}
  \centering
  \begin{tabular}{ccccc|cc|cc}
    \toprule
    \multirow{2}{*}{\textbf{RNN type}} & \multirow{2}{*}{\textbf{Loss}} & \multirow{2}{*}{\textbf{Dense layer neurons}} & \multirow{2}{*}{\textbf{RNN output dimension}} & \multirow{2}{*}{$\mathbf{\gamma}$ }& \multicolumn{2}{c|}{Validation} & \multicolumn{2}{c|}{Test}\\
    \cline{6-9}
    & & & & & \textbf{AUC} & $F_1$ & \textbf{AUC} & $F_1$ \\
    \hline
    LSTM & Weighted focal loss & 200 & 150 & 1 & 0.958 & 0.469 & 0.966 & 0.486 \\
    GRU & Weighted focal loss & 200 & 150 & 2 & 0.947 & 0.425 & 0.972 & 0.468 \\
    LSTM & Weighted cross entropy & 200 & 100 & - & 0.939 & 0.404 & 0.967 & 0.464 \\
    GRU & Weighted cross entropy & 200 & 150 & - & 0.932 & 0.378 & 0.968 & 0.442 \\
    LSTM & Unweighted cross entropy & 200 & 150 & - & 0.899 & 0.727 & 0.948 & 0.794 \\
    GRU & Unweighted cross entropy & 200 & 150 & - & 0.909 & 0.758 & 0.937 & 0.806\\
    GRU NC & Weighted focal loss & 100 & 100 & 1 & 0.922 & 0.322 & 0.909 & 0.324\\
    \bottomrule
  \end{tabular}
  \caption{Results for the best performing models and their best hyperparameters with AUC and $F_1$ scores, evaluated on the validation and test sets. On the bottom row, GRU NC denotes the GRU model with weighted focal loss trained only on time-series data without contextual information.}
  \label{tab:experiment_results}
\end{table*}

\begin{figure*}
     \centering
     \begin{subfigure}[b]{0.3\textwidth}
         \centering
         \includegraphics[width=\textwidth]{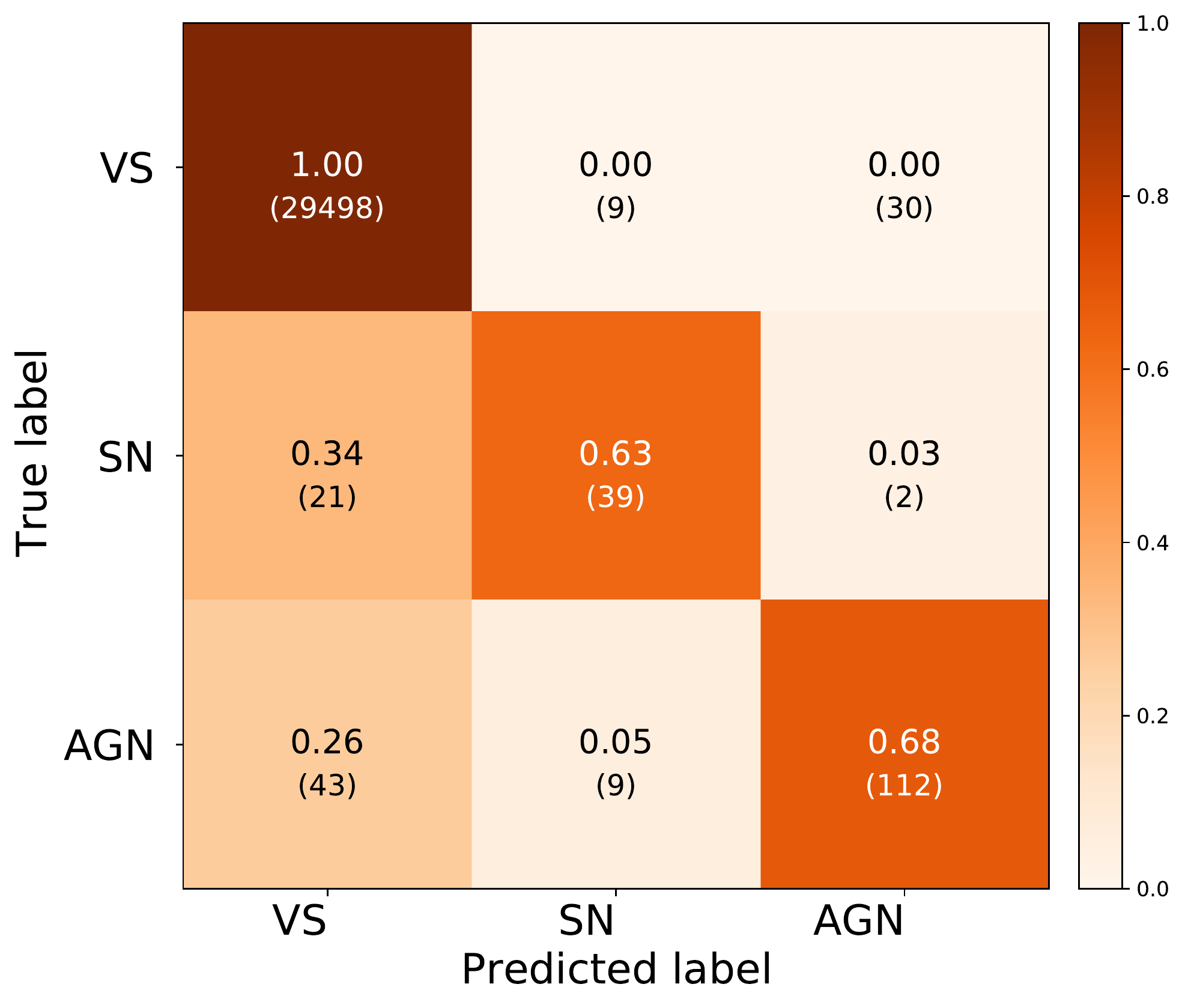}
         \caption{LSTM with unweighted cross entropy}
         \label{fig:lstm_uwce}
     \end{subfigure}
     \hfill
     \begin{subfigure}[b]{0.3\textwidth}
         \centering
         \includegraphics[width=\textwidth]{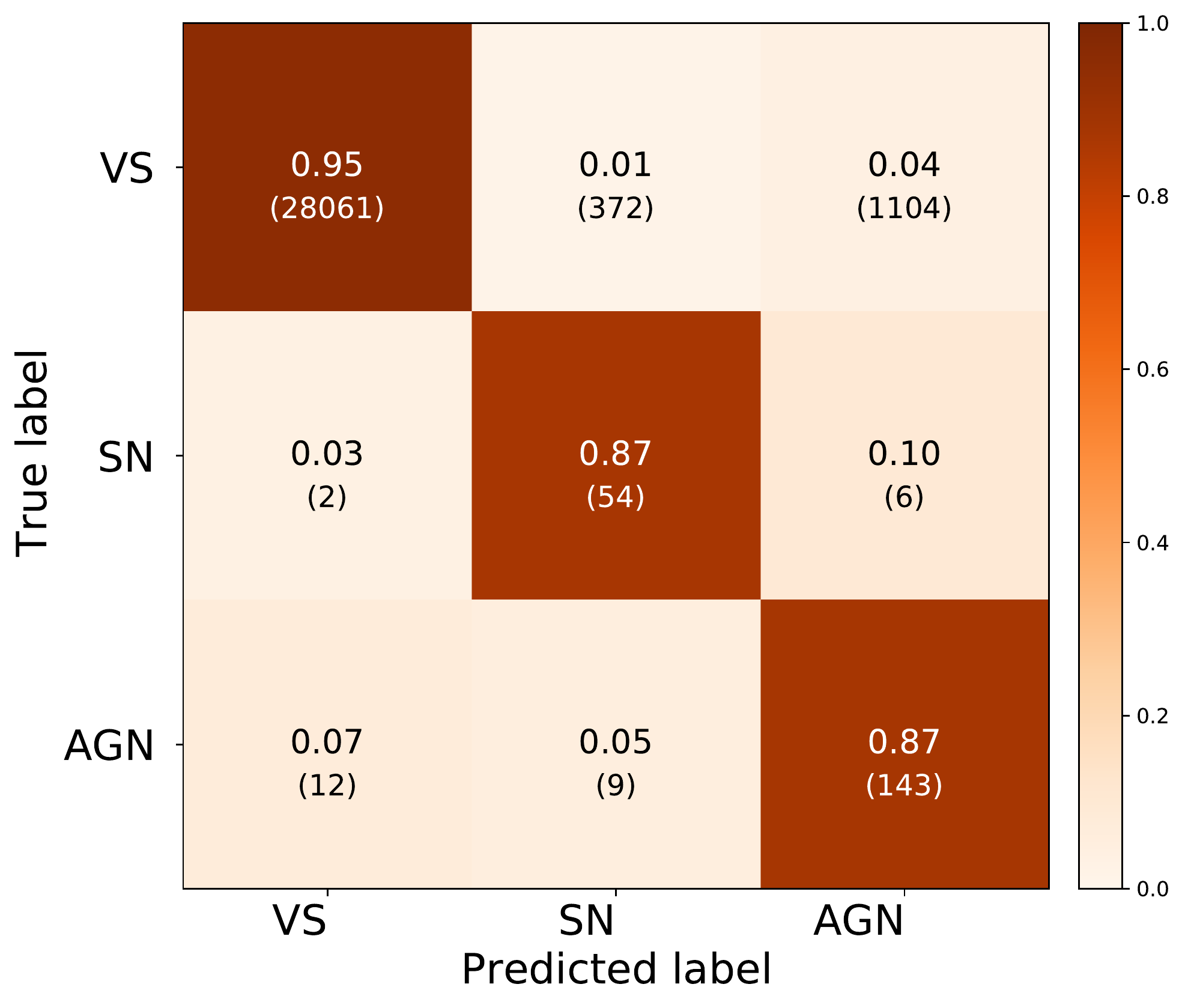}
         \caption{LSTM with weighted cross entropy}
         \label{fig:lstm_wce}
     \end{subfigure}
     \hfill
     \begin{subfigure}[b]{0.3\textwidth}
         \centering
         \includegraphics[width=\textwidth]{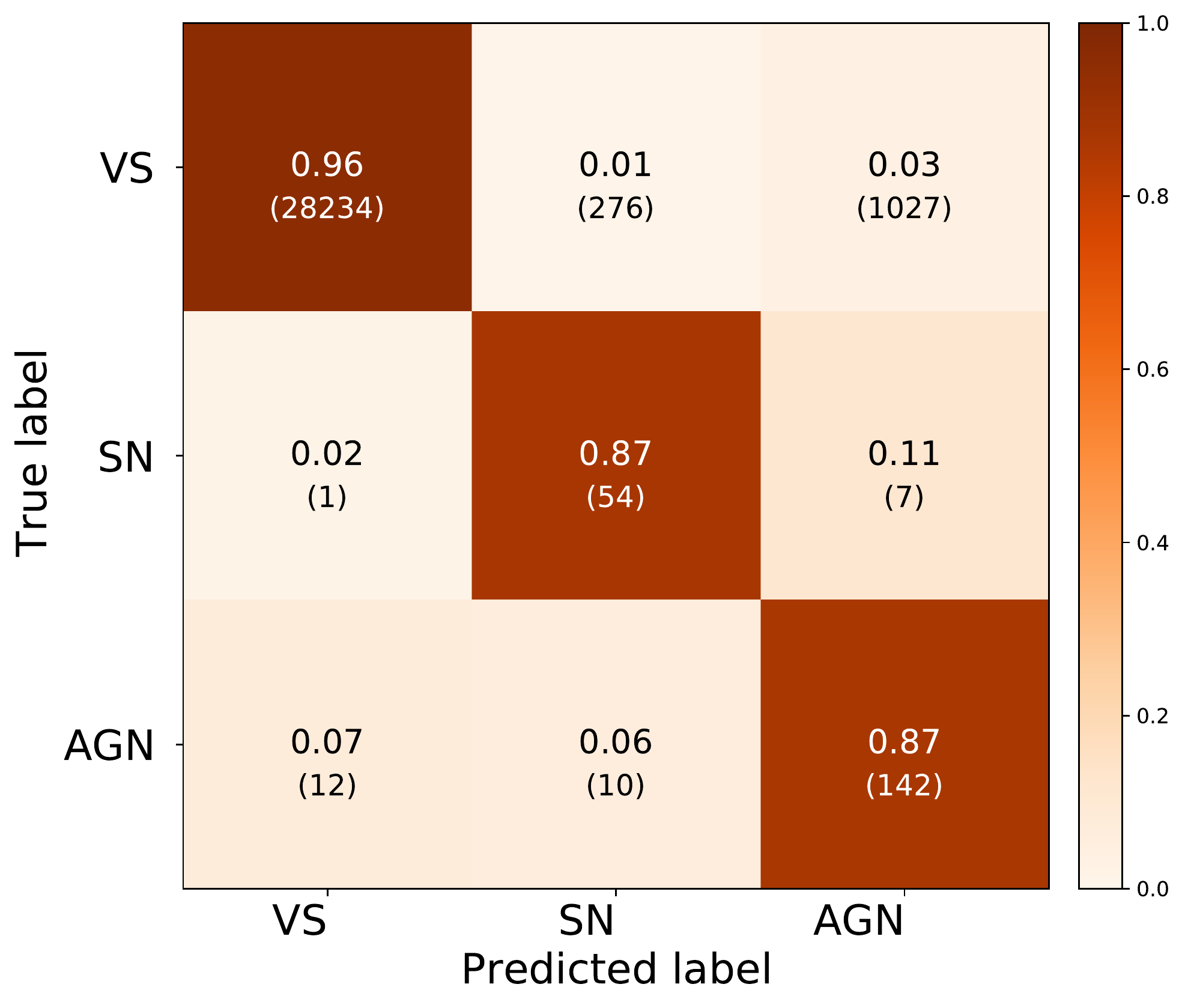}
         \caption{LSTM with weighted focal loss}
         \label{fig:lstm_fl}
     \end{subfigure}
    \newline
         \begin{subfigure}[b]{0.3\textwidth}
         \centering
         \includegraphics[width=\textwidth]{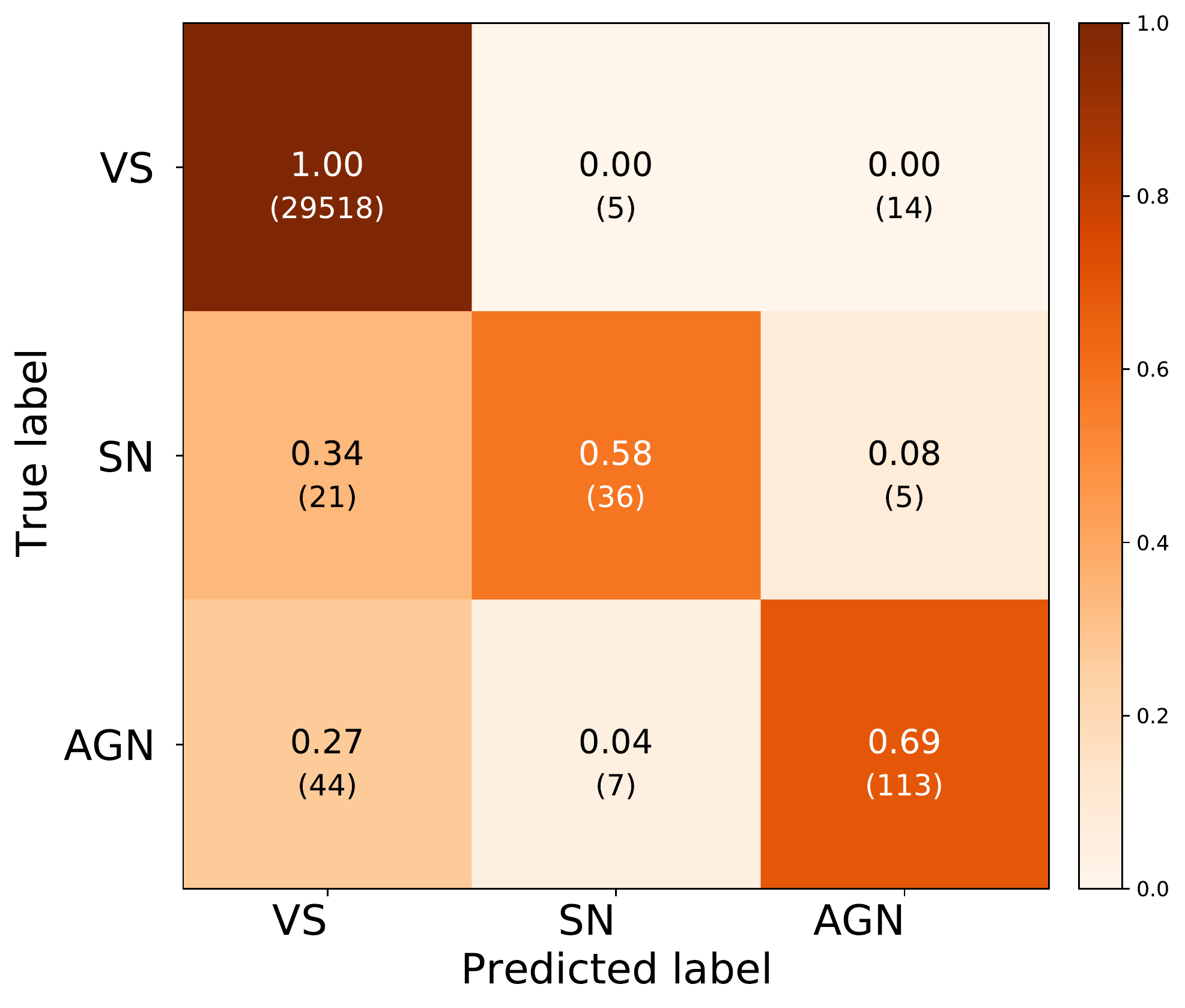}
         \caption{GRU with unweighted cross entropy}
         \label{fig:gru_uwce}
     \end{subfigure}
     \hfill
     \begin{subfigure}[b]{0.3\textwidth}
         \centering
         \includegraphics[width=\textwidth]{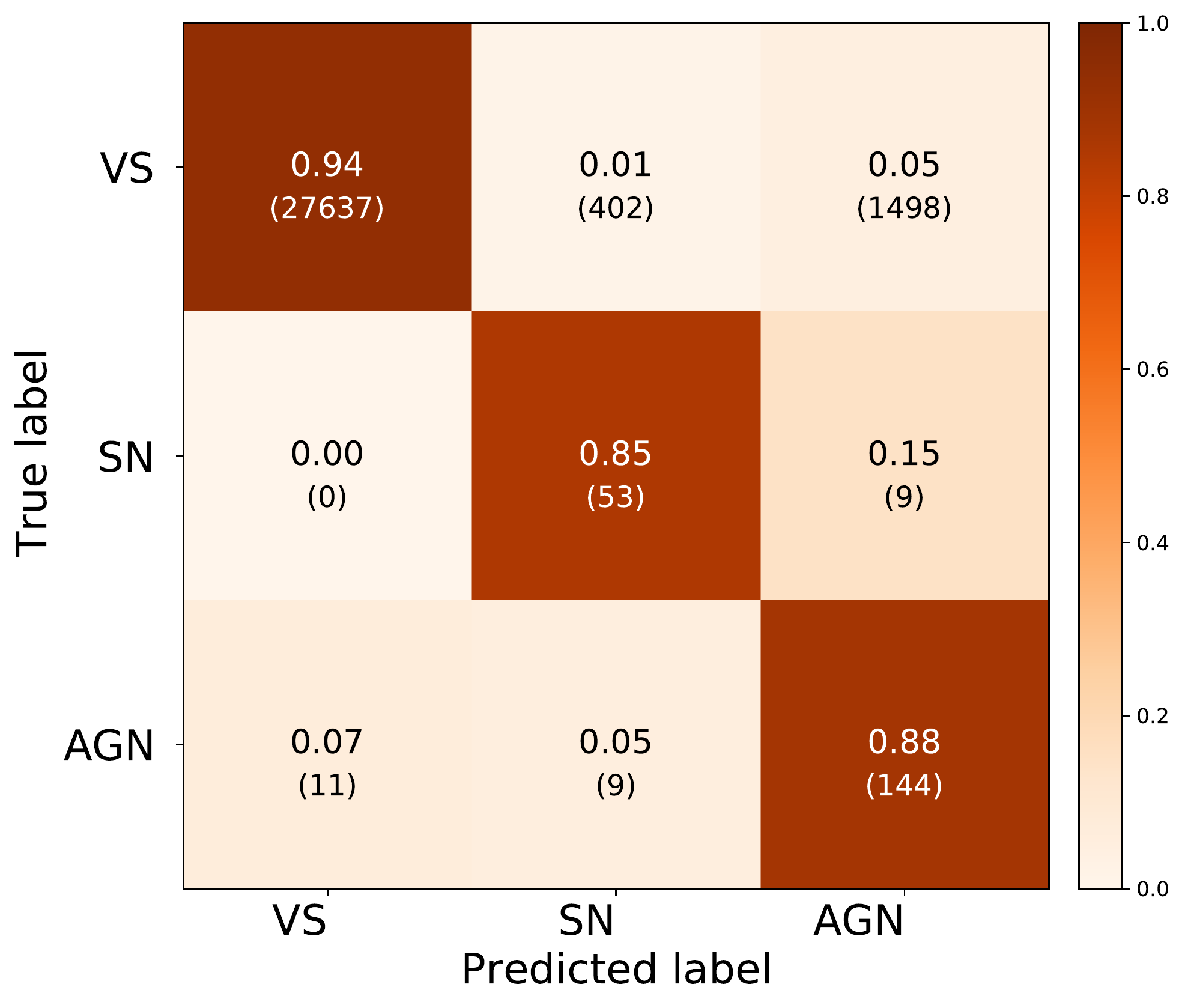}
         \caption{GRU with weighted cross entropy}
         \label{fig:gru_wce}
     \end{subfigure}
     \hfill
     \begin{subfigure}[b]{0.3\textwidth}
         \centering
         \includegraphics[width=\textwidth]{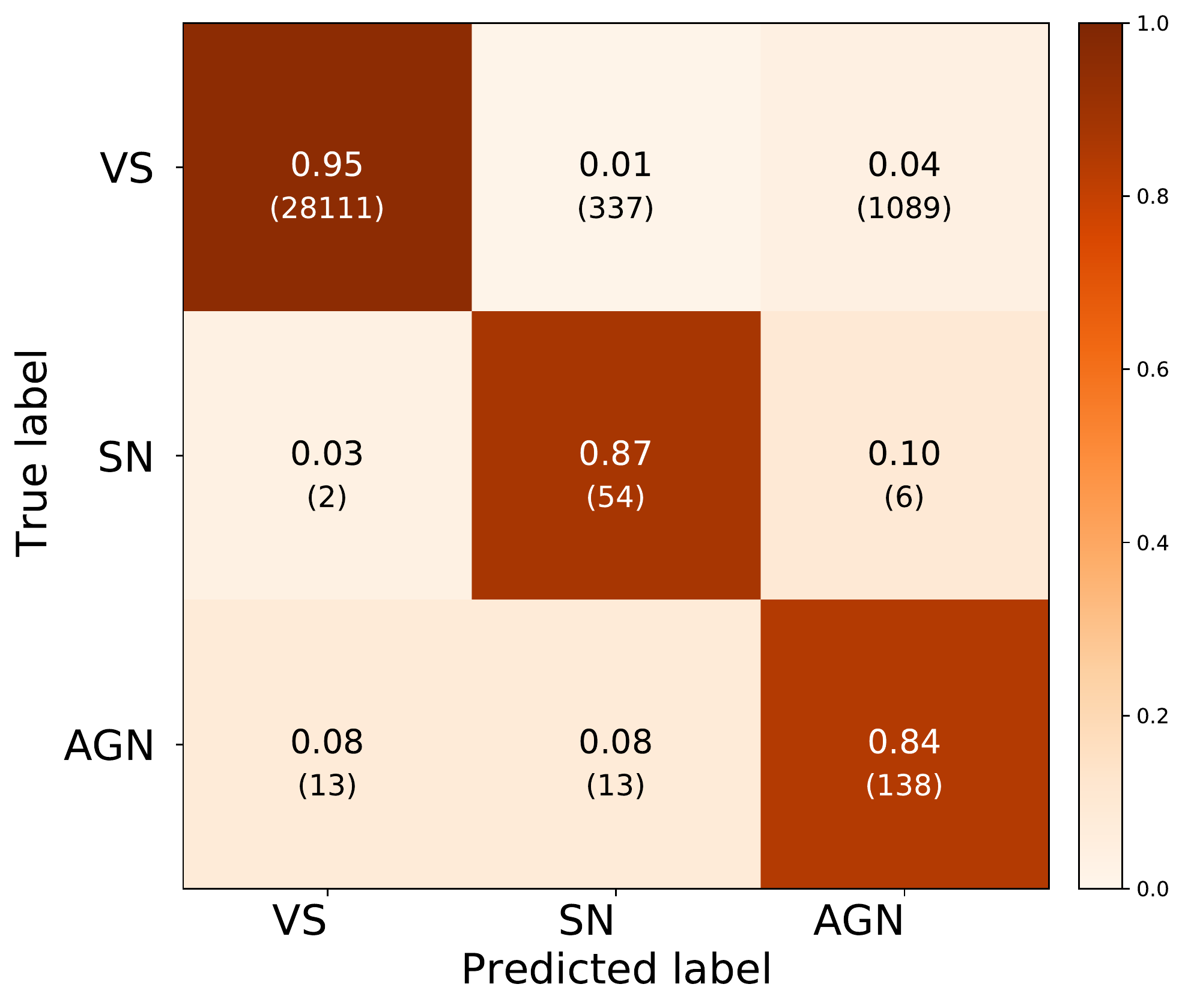}
         \caption{GRU with weighted focal loss}
         \label{fig:gru_fl}
     \end{subfigure}
        \caption{Confusion matrices for the best performing models on test data. The labels on the $x$-axis are the labels predicted by the classifier, and the labels on the $y$-axis are the true labels. Correct predictions are represented by values along the diagonal, incorrect predictions are represented by values in the off-diagonal. The rows of the matrix show the fraction of correct and incorrect predictions for each class, and where incorrect predictions between classes occur. Below the fractions are the number of objects that have been predicted, in parentheses.}
        \label{fig:test_confusion_matrices}
\end{figure*}

\begin{figure}
    \centering
    \includegraphics[width=0.45\textwidth]{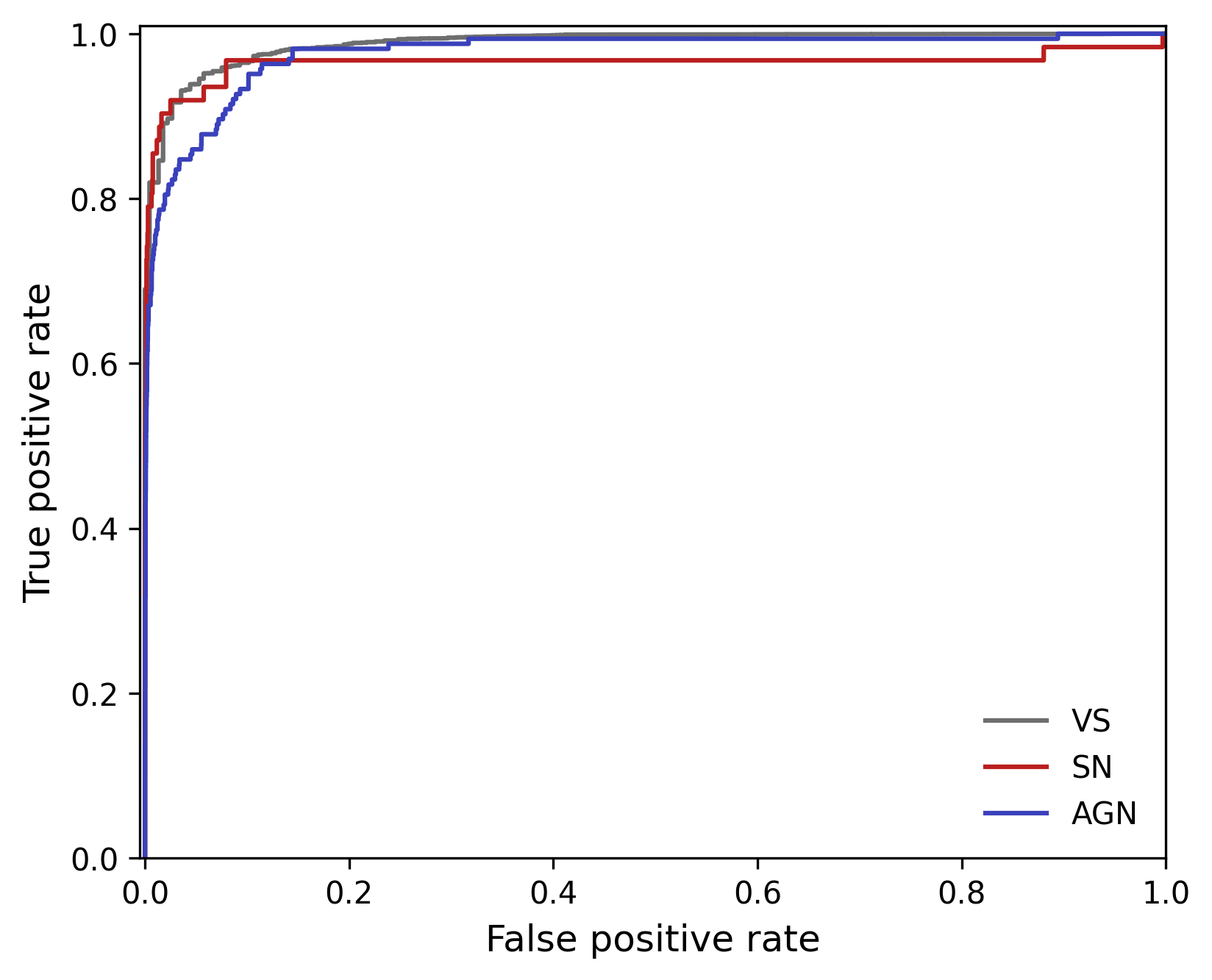}
    \caption{Receiver operating characteristic (ROC) curves for the VS, SN, and AGN classes (grey, red, and blue respectively). ROC curves plot the true positive rate (TPR) against the false positive rate (FPR) for a range of threshold values that dictate whether an object is classified as positive or negative. The curve is obtained by considering a separate binary classification case for each class, treating one class as positive, and the rest as negative.}
    \label{fig:gru_fl_roc_all_epochs}
\end{figure}

\begin{figure}
    \centering
    \includegraphics[scale=0.48]{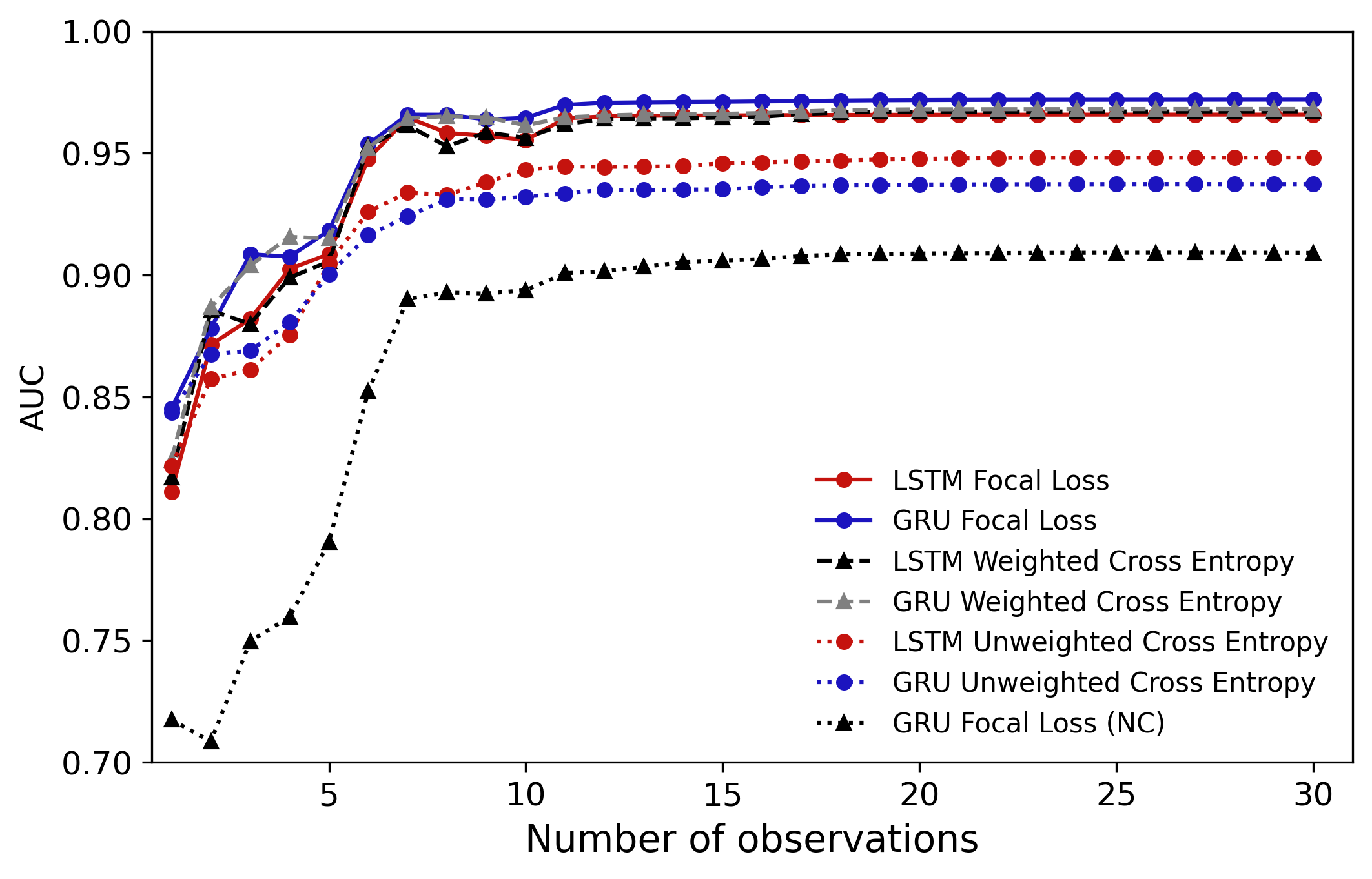}
    \caption{AUC scores evaluated for all models on the test data, plotted as a function of increasing number of light curve observations included in the light curve. The black dotted line with triangular markers shows the AUC scores for the GRU model trained with weighted focal loss without contextual information (labelled GRU Focal Loss (NC)).}
    \label{fig:auc_time}
\end{figure}

\begin{figure*}
     \centering
     \begin{subfigure}[b]{0.3\textwidth}
         \centering
         \includegraphics[width=\textwidth]{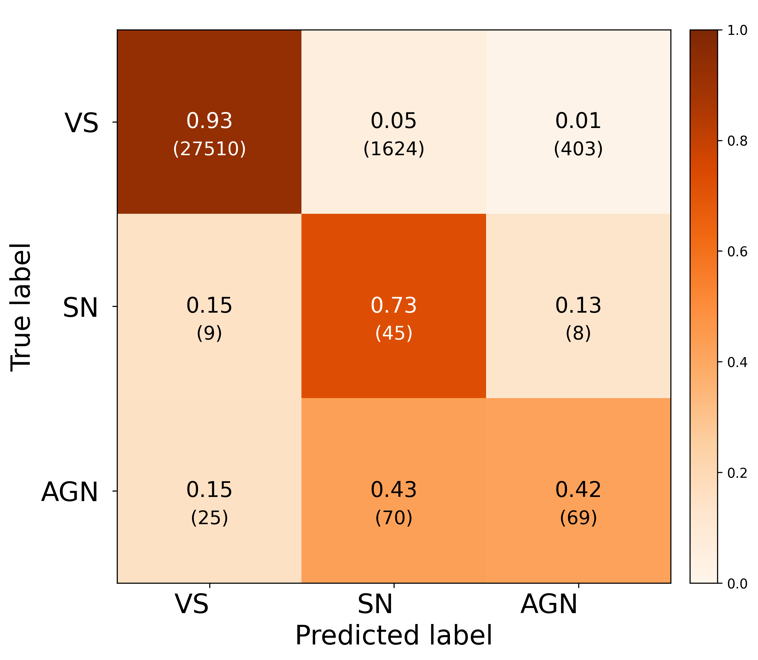}
         \caption{Maximum observations = 1}
         \label{fig:cm_1}
     \end{subfigure}
     \hfill
     \begin{subfigure}[b]{0.3\textwidth}
         \centering
         \includegraphics[width=\textwidth]{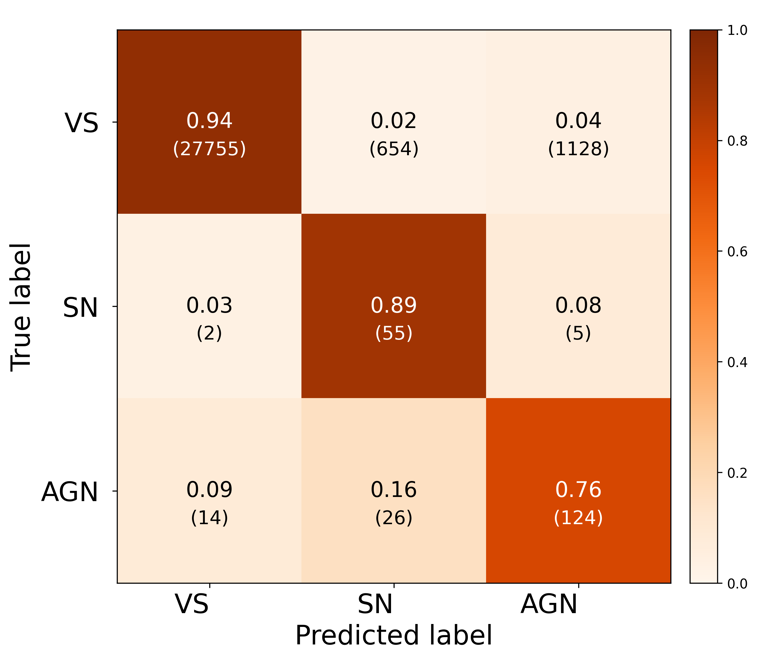}
         \caption{Maximum observations = 6}
         \label{fig:cm_6}
     \end{subfigure}
     \hfill
     \begin{subfigure}[b]{0.3\textwidth}
         \centering
         \includegraphics[width=\textwidth]{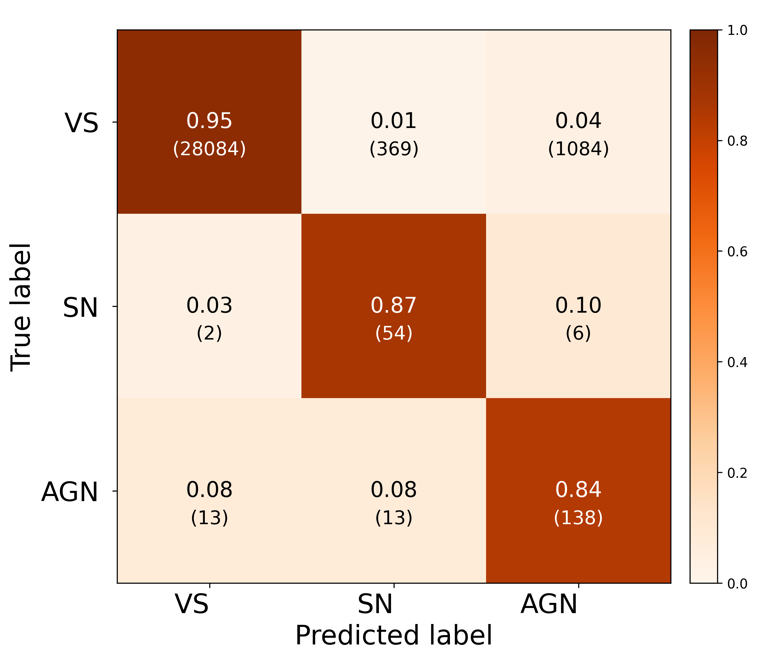}
         \caption{Maximum observations = 20}
         \label{fig:cm_20}
     \end{subfigure}
        \caption{Confusion matrices for the GRU model with weighted focal loss, evaluated with an increasing number light curve observations.}
        \label{fig:cm_time_series}
\end{figure*}

\section{Results}
\label{sec:results}

\subsection{Hyperparameter optimisation}
\label{subsec:hyperparam_ops}

The models are trained with all possible hyperparameter combinations presented in Table \ref{tab:hyperparams}. After training, all models are evaluated on the validation set. The best set of hyperparameters for the six different model configurations are summarised in Table \ref{tab:experiment_results}, with both AUC and the $F_1$ score shown for each model. All models converge within 200 epochs, figure \ref{fig:train_loss} in the appendix illustrates how the loss on the training and validation sets evolve during training.

The weighted cross entropy models show a small increase of $\sim0.03$ in AUC over the unweighted cross entropy models, and the weighted focal loss shows an additional increase of $\sim0.05$ in AUC. In this work we used the unweighted $F_1$ score, which calculates the $F_1$ score for each class and takes the unweighted mean. Although the unweighted cross entropy loss models both have higher $F_1$ scores, the models with weighted loss functions perform better overall across all classes - the $F_1$ score is skewed towards the class with more examples and is not a metric well-suited for imbalanced data. For the weighted focal loss models, the LSTM appears to perform slightly better than the GRU on the validation set.

\subsection{Test set performance}
\label{subsec:test_set_perf}

After the best hyperparameters are determined for all the models, the models are then evaluated on the test set. The test set consists of data that the classifier has not seen during the training phase. Columns 8 and 9 of table \ref{tab:experiment_results} summarises the performance of the six different model configuration on the test set, using all available observations. 

As in the results for the validation set, the performance of the models on the test set indicate that simply adding weights to the loss function to account for class imbalance improves performance. Both the weighted cross entropy and weighted focal loss models show an increase of $\sim0.02$ in AUC over the unweighted cross entropy. The $F_1$ scores are also shown for illustrative purposes; although the models with unweighted cross entropy loss have the highest $F_1$ scores, they have the lowest AUC scores. To better clarify the improvement of weighted loss models over the unweighted cross entropy models, it is prudent to also look at the confusion matrices for each of these models.

Figure \ref{fig:test_confusion_matrices} shows the confusion matrices for all the models evaluated on the test set. Confusion matrices are useful evaluation tools for multi-class classification problems; they visualise how often a classifier makes correct predictions, and where misclassifications between classes occur (Figure \ref{fig:confusion_matrix}). The models with unweighted cross entropy loss perform the worst, and the confusion matrices for these models show the impact of class imbalance on the model. 

Figures \ref{fig:lstm_uwce} and \ref{fig:gru_uwce} show that the unweighted cross entropy loss models have complete recovery for VS, but also misclassify roughly a third of SN and AGN as variable stars. The LSTM model correctly classifies $63\%$ of SN and $68\%$ of AGN, and the GRU model correctly classifies $58\%$ of SN and $69\%$ of AGN. There is relatively little confusion between SN and AGN for the LSTM and GRU models.

Using a weighted cross entropy loss function improves performance on AGN and supernovae, but slightly decreases performance for VS. Figures \ref{fig:lstm_wce} and \ref{fig:gru_wce} show increased recovery for SN and AGN: up to $87\%$ and $85\%$ accuracy for SN with the LSTM and GRU models, respectively, and up to $87\%$ and $88\%$ accuracy for AGN with the LSTM and GRU models, respectively. The accuracy for VS dropped to $95\%$ for the LSTM model and $94\%$ for the GRU model, with misclassifications occurring in both SN and AGN. There is a small amount of confusion ($<15\%$) between SN and AGN for the LSTM and GRU models with weighted cross entropy loss functions.

The models using weighted focal loss show similar performance to the weighted cross entropy models, with minor improvements. Figures \ref{fig:lstm_fl} and \ref{fig:gru_fl} show that both LSTM and GRU models with weighted focal loss are able to achieve $88\%$ accuracy for SN, up to $96\%$ accuracy for VS (with the GRU model achieving $95\%$ accuracy for VS), and up to $87\%$ accuracy for AGN (with the GRU model achieving $84\%$ accuracy for AGN). There is some degree of confusion between SN and AGN, but no more that $11\%$ of examples from these classes are misclassified as the other. Only up to $3\%$ of examples from SN are classified as VS, and up to $8\%$ of AGN are classified as VS. This is a significant improvement over the models with unweighted cross entropy, where up to a third of objects from the minority classes (SN, AGN) are misclassified as the majority class (VS).

\begin{table}
    \centering
    \begin{tabular}{l|c|c|c}
        \toprule
         \textbf{Class} & \textbf{Threshold} & \textbf{TPR} & \textbf{FPR}\\
         \hline
         VS & 0.7 & 86.2\% (25,465) & 1.8\% (4) \\
         SN & 0.7 & 74.2\% (46) & 0.2\% (57) \\
         AGN & 0.7 & 72.6\% (119) & 0.7\% (213) \\
         \bottomrule
    \end{tabular}
    \caption{The true positive rate (TPR) and false positive rate (FPR) for each class, evaluated on the test set at a threshold for the GRU model trained with focal loss. The positive and negative predictions are obtained by treating each class as positive, and the other two as negative, creating a binary classification problem for each of the three classes. The number of true positive predictions and false positive predictions are shown in parentheses with the TPR and FPR values.}
    \label{tab:prediction_rates}
\end{table}

Looking at the number of predictions made for each entry in the confusion matrix for the GRU model trained with focal loss (Figure \ref{fig:gru_fl}), it can be seen that the number of VS predicted as SN and AGN is greater than the number of actual SN and AGN in the test set. The classifier is designed to produce prediction probabilities, and the predicted class is simply selected by choosing the class prediction that has the highest probability. We can examine how varying the threshold value for the class probability can refine the classification results. By breaking down the three class problem into three binary classification problems, where for each case the positive class is one of VS, SN, or AGN, and the negative class are the other two, we can calculate the true positive rate (TPR) and false positive rate (FPR) for each case. The probability of a negative prediction is simply given as the sum of the probability of the two non-positive classes. We use the predictions given by the GRU model trained with focal loss since it has the highest AUC score on the test set, and compute the TPR and FPR., and produce a ROC curve for each class (Figure \ref{fig:gru_fl_roc_all_epochs}). The ROC curves for all the classes reflects the high AUC score of the GRU model trained with weighted focal loss, covering most of the TPR-FPR space and reaching the top left-hand corner (high TPR at low FPR).

We can examine how varying the threshold can reduce contamination, that is, to reduce the number of false positives in each class. By selecting a 'cut-off' threshold, any objects that have a prediction probability below the threshold can be regarded as negative, and those with a prediction probability above the threshold can be regarded as positive. Table \ref{tab:prediction_rates} shows the TPR and FPR along with the number of true positive and false positive predictions for each class by using a threshold value of 0.7. By selecting a cut-off threshold, the number of false positive predictions for all classes is reduced. At a threshold value of 0.7, the TPR of SN and AGN is $>70\%$ at a FPR of $<1\%$. The number of false positives for SN and AGN is still significant, and we note that reducing contamination on minority classes in an imbalanced data setting remains a challenge.

\subsection{Time-dependent performance}
\label{subsec:time-results}

With an RNN architecture, it is possible to take in sequential inputs of different lengths. Hence, it is possible to evaluate the classifiers performance by varying the number of light curve observations used. To do this, the time-series input matrix can be formatted so that it contains only the first $n$ observations, and the remaining values are padded. The models are then evaluated on the test data, using an increasing number of observations from $n=1$ to a maximum of $n=30$. 

Figure \ref{fig:auc_time} shows how the AUC of all models vary as the number of observations included in the light curves are increased. In this case, the number of observations refers to the maximum number of observations that are included. If the maximum number of observations is $m$, then a light curve with fewer than $m$ observations will have all its observations included. If a light curve has more than $m$ observations, then only the first $m$ observations are used.

With just one light curve observation, the unweighted cross entropy models, weighted cross entropy models, and the LSTM model with weighted focal loss achieve AUC scores of $\sim0.82$. The GRU model with weighted focal loss achieves the highest AUC score with a single light curve observation with 0.84. As more observations are included, all models show an increase in AUC until around ten observations, after which the AUC scores maintain a constant value. Since a majority of the light curves in the data set only have up to ten observations (see Figure \ref{fig:lc_statistics}), it is not surprising that model performance remains constant after a maximum of ten light curve observations are included. The final values of the AUC for all models are shown in column 8 of table \ref{tab:experiment_results}. 

We perform some additional analysis on how the GRU model with weighted focal loss performs over time, choosing the aforementioned model since it has the highest AUC score on the test set. Figure \ref{fig:cm_time_series} shows the confusion matrices for the GRU model with weighted focal loss evaluated with different numbers of light curve observations. With one observation, the model is already able to separate out VS from other objects to a good degree of accuracy, achieving $93\%$ accuracy for VS. At one epoch of observation, the model achieves $73\%$ accuracy for SN, and $42\%$ accuracy for AGN, with $43\%$ of AGN being misclassified as SN. Other than AGN being misclassified as supernovae, there is some degree of confusion between all classes: $13\%$ of SN are misclassified as AGN, and $15\%$ of SN and AGN are misclassified as variable stars. 

With up to six light curve observations, the accuracy for SN and AGN improves. $89\%$ of supernovae and $76\%$ of AGN are correctly classified, with some misclassifications between the two ($<16\%$), and few SN and AGN being misclassified as variable stars. At a maximum of twenty observations, the model reaches the maximum performance. The confusion matrix for the GRU model with weighted focal loss at twenty epochs in figure \ref{fig:cm_20} appears similar to the confusion matrix evaluated with all epochs of observations as in figure \ref{fig:gru_fl}. The SN accuracy drops from $89\%$ at six observations to $87\%$ at twenty observations, suggesting some confusion as more observations are included and light curves appearing similar to each other at longer timescales.

\subsection{Importance of contextual information with t-SNE}
\label{subsec:contextual_information}

Imaging surveys such as GOTO will be able to provide contextual information for newly discovered objects (for example, cross matching to galaxy catalogs) in addition to photometric data. We now discuss the importance of contextual information for the model in learning to differentiate between objects from different classes. To investigate the impact of contextual information, we train a grid of GRU models with weighted focal loss using the hyperparameters in Table \ref{tab:hyperparams} on the same data as the other models, but without additional contextual information. The inputs to these models are just the time-series data from the light curves $\mathbf{X_T}$. We identify the best performing model by selecting the model that achieves the highest AUC score on validation set.

\subsubsection{Performance without contextual information}
\label{subsubsec:nc_performance}

\begin{figure*}
     \centering
     \begin{subfigure}[b]{0.3\textwidth}
         \centering
         \includegraphics[width=\textwidth]{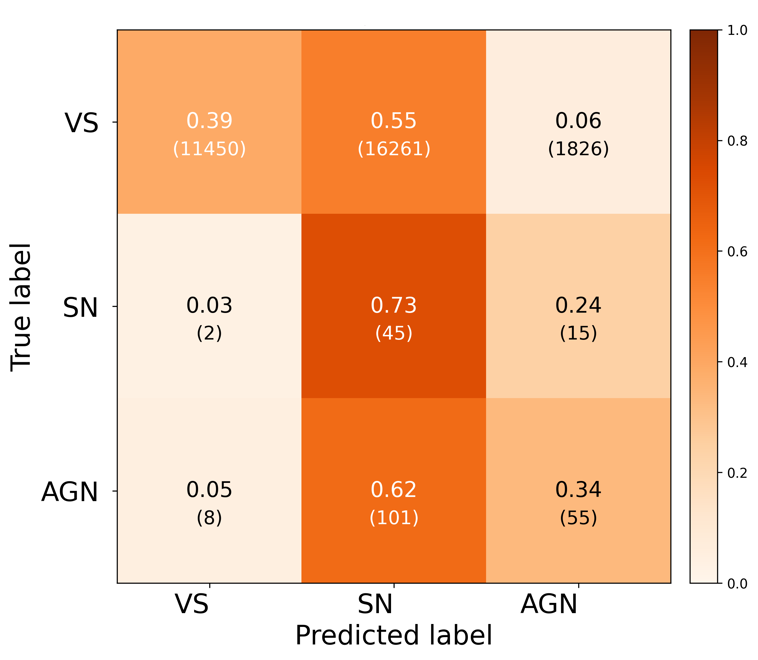}
         \caption{Maximum observations = 1}
         \label{fig:cm_1_ts}
     \end{subfigure}
     \hfill
     \begin{subfigure}[b]{0.3\textwidth}
         \centering
         \includegraphics[width=\textwidth]{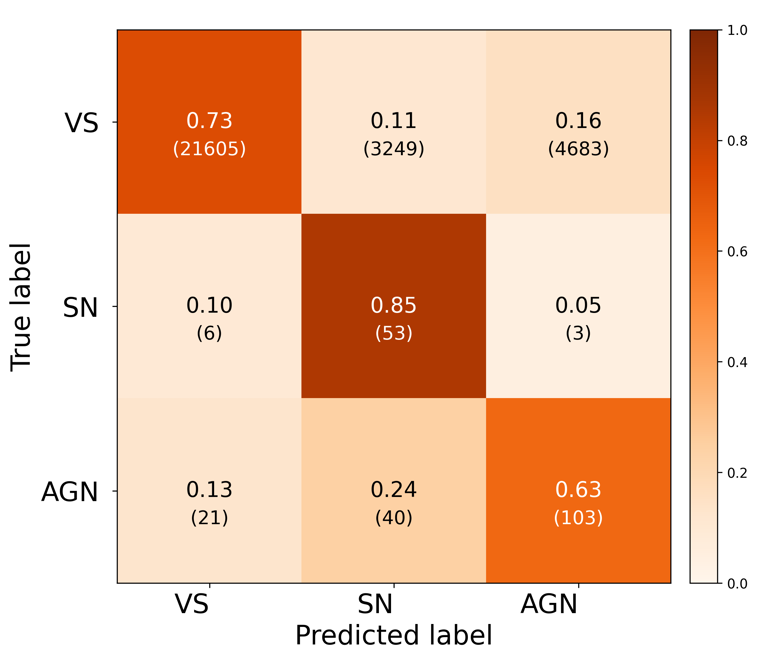}
         \caption{Maximum observations = 6}
         \label{fig:cm_6_ts}
     \end{subfigure}
     \hfill
     \begin{subfigure}[b]{0.3\textwidth}
         \centering
         \includegraphics[width=\textwidth]{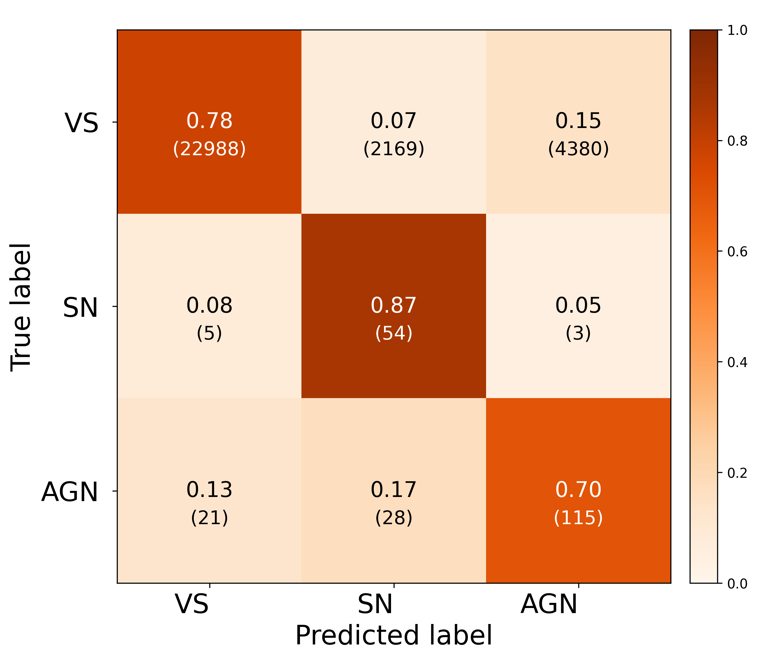}
         \caption{Maximum observations = 30}
         \label{fig:cm_20_ts}
     \end{subfigure}
        \caption{Confusion matrices for the GRU model with weighted focal loss trained only on time-series data, evaluated with an increasing number light curve observations.}
        \label{fig:cm_time_series_timeonly}
\end{figure*}

\begin{figure}
     \centering
    \includegraphics[width=0.45\textwidth]{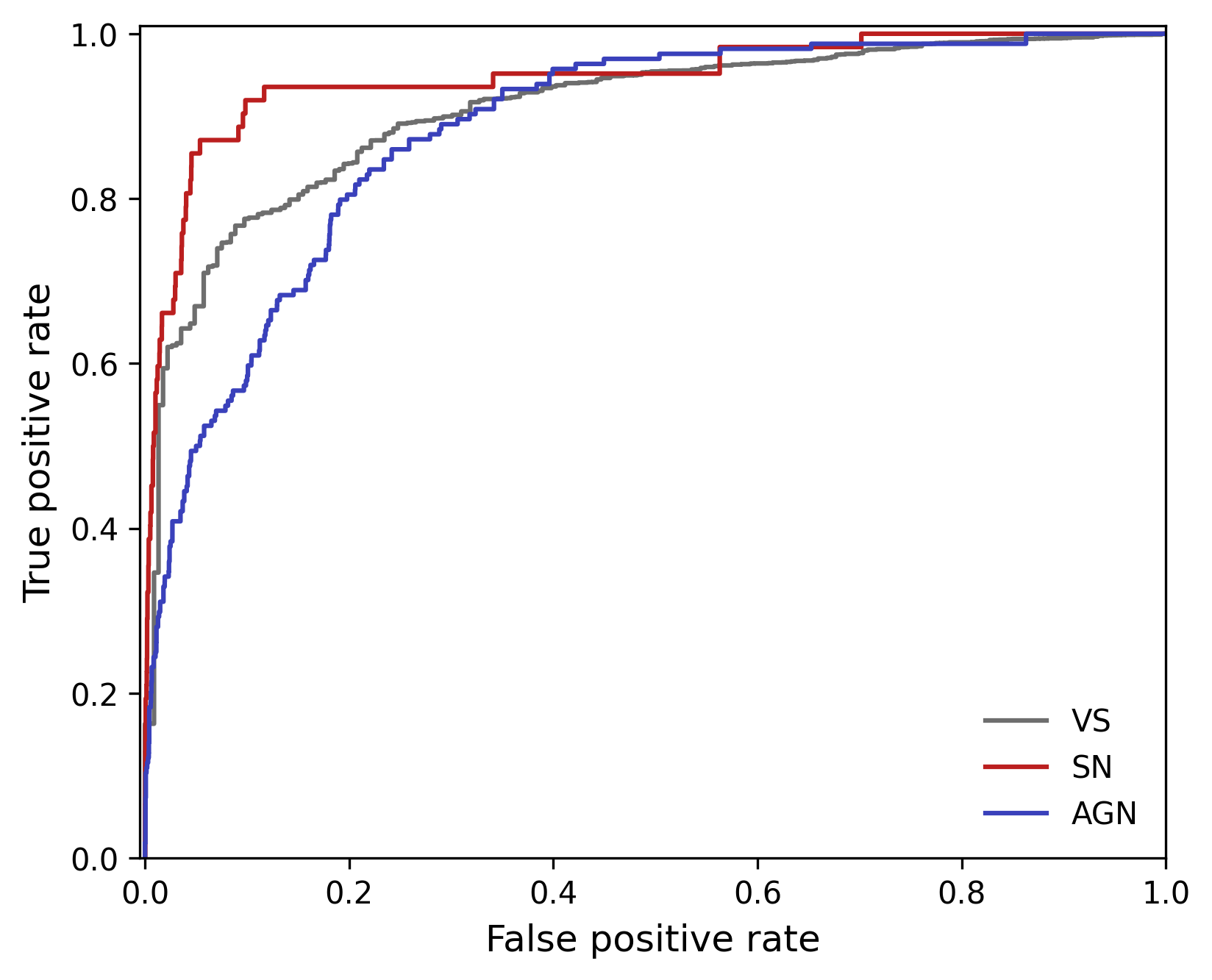}
    \caption{ROC curves for the VS, SN, and AGN classes (grey, red, and blue respectively) for the GRU model with weighted focal loss, trained only on time-series data.}
    \label{fig:roc_gru_fl_timeonly}
\end{figure}

The hyperparameters, AUC and $F_{1}$ scores of the best performing GRU model with weighted focal loss trained only on time-series data is shown in the bottom row of Table \ref{tab:experiment_results}. The model achieves an AUC score of 0.922 on the validation set, which is higher than the validation AUC scores of the unweighted cross entropy models.

We follow the same process as in section \ref{subsec:test_set_perf} and evaluate this model on the test set. The model achieves an AUC score of 0.902 on the test set, which is the lowest AUC score out of all models. In figure \ref{fig:auc_time}, we plot how the AUC score evolves as more light curve observations are included. At one observation, the model achieves an AUC score of $0.717$ and then increases up until the six observations after which is starts to maintain a constant value.

Figure \ref{fig:cm_time_series_timeonly} shows the confusion matrices evaluated on the test set using an increasing number of maximum observations. With just one light curve observation, the model performs worse than the GRU model with weighted focal loss trained with contextual information. The model achieves an accuracy of $39\%$ for VS and incorrectly classifying $55\%$ of VS as SN, and $34\%$ accuracy for AGN and incorrectly classifying $62\%$ of AGN as SN. SN accuracy for the model is similar to the models trained with weighted focal loss, with an accuracy of $73\%$. When using all available light curve observations, the model trained only on time series data achieves a similar accuracy for SN as the weighted focal loss models at $87\%$, but lower accuracy for VS at $78\%$ and AGN at $70\%$. There is a slightly higher degree of misclassification between classes compared to the models with weighted loss functions trained with contextual information.

Figure \ref{fig:roc_gru_fl_timeonly} shows the ROC curve for the model trained only on time-series data (using all available light curve observations), by separating the three-class problem into three separate binary classification problems. Compared to figure \ref{fig:gru_fl_roc_all_epochs}, the ROC curves show that the model does not perform as well when contextual information is excluded.

Overall, the model trained only with time-series data performs worse than its counterpart trained with contextual information, but is able to achieve comparable accuracy for SN. This suggests that the model is able to extract information from the light curves that allows for good separation of SN from the other classes. We expand on this in the following analysis.

\subsubsection{t-Distributed Stochastic Neighbor Embedding representation}
\label{subsubsec:tsne_analysis}

\begin{figure*}
    \centering
        \begin{subfigure}[b]{0.47\textwidth}
            \centering
            \includegraphics[width=\textwidth]{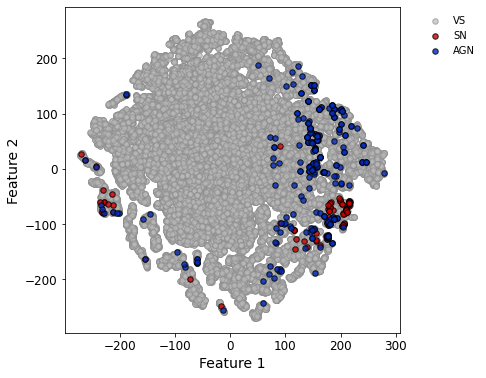}
            \caption{t-SNE representation of model output after the final GRU layer, with no contextual information.}
            \label{fig:gru_timeonly_tsne}
        \end{subfigure}
        \hfill
        \begin{subfigure}[b]{0.47\textwidth}
            \centering
            \includegraphics[width=\textwidth]{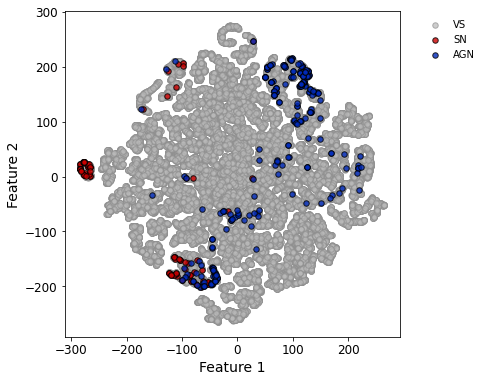}
            \caption{t-SNE representation of model output after the dense layer before the output layer, with no contextual information.}
            \label{fig:dense_timeonly_tsne}
        \end{subfigure}
    \newline
        \begin{subfigure}[b]{0.47\textwidth}
            \centering
            \includegraphics[width=\textwidth]{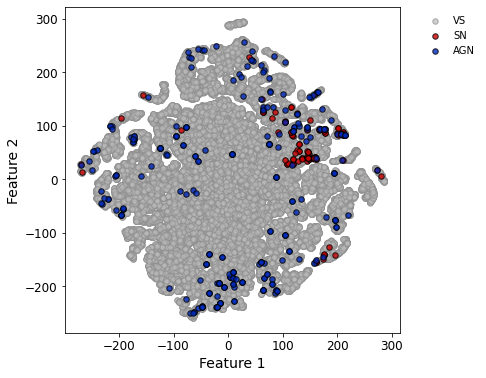}
            \caption{t-SNE representation of model output after the final GRU layer, with contextual information.}
            \label{fig:gru_tsne}
        \end{subfigure}
        \hfill
        \begin{subfigure}[b]{0.47\textwidth}
            \centering
            \includegraphics[width=\textwidth]{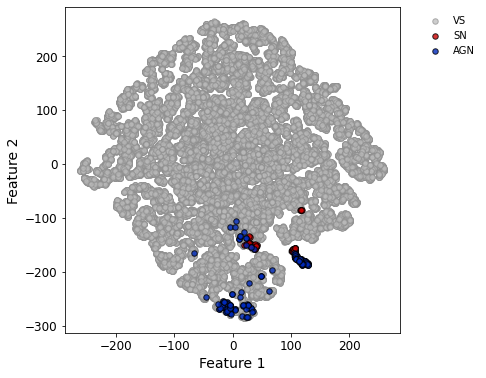}
            \caption{t-SNE representation of model output after the dense layer before the output layer, with contextual information.}
            \label{fig:dense_tsne}
        \end{subfigure}

    \caption{t-SNE representation for the network outputs at different stages. Each datapoint corresponds to a single object in the training set; grey points are VS, red points are SN, and blue points are AGN.}
    \label{fig:tsne}
\end{figure*}

We use t-Distributed Stochastic Neighbor Embedding (t-SNE) to represent how the model transforms the input data at different layers of the network. t-SNE is a data visualization technique used to map a high-dimensional data set into a low-dimensional data set that can be visualised in a two or three dimensional scatter plot \citep{2008vandermaaten}. We provide a summary of how t-SNE makes a low-dimensional visualization of a high-dimensional data set in the appendix.

t-SNE has been used to visualise class separability in supernovae classification with ML algorithms \citep{2016lochner}. In this analysis, we use t-SNE to visualise how class separability changes at different layers in the network. We consider two models: the GRU model with weighted focal loss trained with contextual information and the same model trained with only time-series data. 

We take the output of the model at two points: the output after the final GRU layer, where the model only considers time-series information from the light curve, and the output after the final dense layer before the output layer. In the model trained with contextual information, the output of the final dense layer will encode the contextual information introduced after the final GRU layer. The model is fed the training data as input, and t-SNE is used to produce a low-dimension visualization of the high-dimensional intermediate outputs. Figure \ref{fig:tsne} shows the two-dimensional mapping of the model outputs at the final GRU layer and the final dense layer, for both models. The \texttt{scikit-learn} Python package \citep{scikit-learn} implementation of t-SNE is used in this analysis.

Looking at the t-SNE representation of the outputs for the model trained without contextual information, there appears to be no clear clustering in the representation for the output after the GRU layer in figure \ref{fig:gru_timeonly_tsne}. The majority of AGN occupy the right side of the plot, the majority of SN are sparsely clustered in the bottom-right region, and few AGN and SN occupy the left side. In figure \ref{fig:dense_timeonly_tsne}, the t-SNE representation of the output after the final dense layer shows some coherent clustering of SN and AGN. AGN cluster around the top-right and bottom-left region, with some spread out in the middle. There is a compact structuring of SN on the left side, and a cluster of SN in the bottom-left near the AGN. There are a few AGN and SN in the top-left region of the plot.

From figure \ref{fig:gru_tsne} for the output after the final GRU layer for the model trained with contextual information, there appears to be no clear clusters of objects from the same class. AGN are spread out across the plot, and there is some clustering of SN in the top-right region. Looking at figure \ref{fig:dense_tsne} after including contextual information, the clustering of objects from the same class becomes more apparent. There appears to be a compact and distinct cluster of AGN near the bottom of the plot, and a tight clustering of SN along with AGN in the bottom-right region. There is some overlap between SN and AGN, indicating where some of the misclassifications are occurring.

We note that the VS class is a very broad class, containing a multitude of different variable objects with distinct subclassifications. In figure \ref{fig:dense_tsne}, there seems to be a few coherent structures which may indicate where variable objects from the same or similar classes are located.

From Figure \ref{fig:tsne}, it is clear that the incorporation of contextual information into the classifier provides useful information that allows the model to learn better class separability. Using t-SNE in this way is a method to provide an approximate measure of `feature importance' in deep learning models, which can be challenging compared to deriving feature importance in feature-based ML algorithms such as random forests which utilize hand-made features. 

It should be noted that t-SNE is primarily a \textit{data visualization} tool, and here it is used to visualise class separability learnt by the model at different stages. It is possible to explore the hyperparameter space when generating t-SNE plots, and use a number of diagnostics to assess the `quality' of the dimensionality reduction (such as comparing distances between points in high and low-dimensional space) and identify any correlating features within the high-dimensional data set \citep{2020chatzimparmpas}. However, further interpretation of the t-SNE plots is beyond the scope of this work.

\section{Discussion}
\label{sec:discussion}

Here, we provide a discussion on the task of object classification in the context of the GOTO survey, how we handle class imbalance in this work, and how the use of contextual information in addition to time-series data from light curves can help classifiers produce more robust classifications.

\subsection{Classification strategy for GOTO}
\label{subsec:classification_strat}

In this work, the data set of 99,201 labelled objects are split into three broad classes: variable stars, supernovae, and AGN. There are over 98,000 variable star light curves in the data set, and more than 350 unique class labels provided in the AAVSO catalog \citep{2006watson}. A number of these labels include a mixture of classes, where there is no certainty to which class an object belongs to. Grouping all variable type objects into a single super-class makes the classification task simpler for the model, and allows for a reliable early-time classifications. We also assume that the classifications provided in these catalogs are the ground-truth. To create an extended and more robust labelled dataset, additional catalogs can be used to provide more labelled examples, and also verify the labels provided for already known objects.

The data set used to train the classifier was obtained from observations conducted in the GOTO all-sky survey mode. Given that the light curves in the data set can be quite sparse, and are only in a single filter, this may have presented a challenge for classification into multiple object subtypes. A more refined classifier can be trained to classify between subtypes of objects. An approach for a more refined classification could be to take a hierarchical classification scheme as in \cite{2020Hosenie}, where once a subset of objects have been classified into broad types, another classifier is used to further classify the subset into more specific classes.

Targeted sub-sky surveys that focus on observing a smaller area of sky and local galaxy clusters in search of transients present a chance to conduct observations at higher cadences and in multiple filters. Additional colour information and more uniform light curve sampling can provide data that can be used to train models to produce better and more refined classifications into object subtypes. Models trained on light curves in multiple filters have been shown to be able to differentiate between different types of transients and supernovae subtypes \citep{2019Villar, 2019muthukrishna}.

One of the objectives of this work was to highlight the usefulness of RNNs for real-time object classification in surveys. Work is currently ongoing to implement this classifier in the GOTO discovery pipeline, where new objects observed by GOTO are classified and the classification is returned to the GOTO Marshall (Lyman et al. in prep.), where it can be displayed in a web interface for users to see. The classifier returns class probabilities that give a measure of how confident the classifier thinks an object belongs to certain class. This information can be used to decide to trigger additional follow-up observations. With the RNN architecture, it is possible to update the classification probability as new observations of an object are made.

\subsection{Handling class imbalance in deep neural network architectures}
\label{subsec:class_imbalance_in_dnns}

The dataset used to train the RNN classifier for GOTO light curves presented a class imbalance problem. Other works dealing with class imbalance have utilised data augmentation methods for classification with feature based machine learning algorithms \citep{2020Hosenie, 2019Villar, 2019Boone, 2018revsbech} and achieved good performance. Here, we provide an alternative to data augmentation with a RNN classifier, and use an algorithm-level approach to dealing with imbalance, where a weighted cross entropy loss function and a weighted focal loss function are used to account for imbalanced class distribution in the training set. We train two types of RNN architectures, the LSTM and GRU, with three different loss functions: an unweighted cross entropy, a weighted cross entropy, and a weighted focal loss.

Weighting the cross entropy loss function shows an improvement over the unweighted cross entropy, going from an AUC of 0.948 to 0.968 for the best unweighted cross entropy and weighted cross entropy loss models when evaluated on the test set, respectively. The confusion matrices show that the degree of confusion between the majority and minority classes is reduced by simply weighting the cross-entropy loss function.

The focal loss models perform similar to the weighted cross entropy loss models. The AUC for the best focal loss model was 0.972, evaluated on the test set. For the focal loss models, the GRU model achieves a higher AUC score. We have shown that by using an appropriate loss function to account for imbalanced data, it is possible to achieve good real-time classification of transient and variable sources, without having to artificially augment the training data.

\cite{2016Krawczyk} note that even if a data set is heavily imbalanced, if the classes are well represented and come from non-overlapping class distributions, it is still possible to achieve good classifications. Future surveys may benefit from obtaining spectroscopy of a wider diversity of targets, and not just based on good signal-to-noise ratios. Having good spectroscopic coverage of sources over a range of magnitudes can help data augmentation efforts to create representative training sets for supernovae classification \cite{2020carrick}. Active learning, a class of ML algorithms used to optimise a labelling strategy for unlabelled data, has been used to select spectroscopic follow-up for objects that would give the best improvement to a learning model in supernova classification \citep{2019ishida_activelearning}. Data augmentation methods could be combined with the approach used in this work for an improvement in classification performance in a class imbalanced setting.

\subsection{Contextual information}
\label{subsec:end_contextual_information}

We trained a GRU model with weighted focal loss to classify light curves using only time-series data without additional contextual information, to investigate the impact of contextual information on classification. The model trained without contextual information achieved an AUC score of 0.902 on the test set, which is the lowest AUC score on the test set out of all models. Without contextual information, the model has an overall worse performance compared to the models with weighted focal loss functions that incorporate contextual information.

Using t-SNE to visualise the intermediate outputs of the GRU models trained with and without contextual information, it can be seen that using contextual information such as location in the sky and distance to the nearest galaxy allows the model to learn better class separability compared to just using information from the light curve. In principle, the model architecture used in this work where contextual information is ingested into the model separately to the time-series information from the light curve can be applied to any survey. 

Only three values were used as the contextual information input into all the models: the galactic latitude and longitude, and the distance to the nearest galaxy from the object. It is possible to include additional information, such as additional information about nearby galaxies (colour and metallicity), redshift, and galactic extinction. 

In the era of large sky surveys such as ZTF and the Vera Rubin LSST, and the availability of multiple alert brokers \citep{2020fink, 2020alerce, 2019lasair, 2018narayan}, additional information on newly discovered objects other than photometric data should be leveraged to provide accurate real-time classification of objects. Being able to classify explosive transients early in their light curve evolution allows for follow-up in the early stages of evolution that can provide constraints on explosion mechanisms and progenitor environments \citep{2018zhang, 2016khazov}. 

\section{Conclusion}
\label{sec:conclusion}

In this paper we present a recurrent neural network classifier to classify objects observed with the GOTO survey using their light curves, and additional contextual information such as on-sky position and distance to the nearest galaxy obtained by cross-matching with a catalog. We create a labelled dataset from the GOTO data, and split the dataset into three classes: variable stars (VS), supernovae (SN), and active galactic nuclei (AGN). The dataset is imbalanced, with 99\% of labelled objects belonging to the variable star class. We adopt weighted cross entropy and focal loss functions to account for the imbalance, and reduce the model bias towards the majority class. The weighted loss functions improves overall classification performance over the standard approach of an unweighted cross entropy loss function with deep neural network classifiers. We also train a model without contextual information and only time-series data, and find that it performs worse than the model with the same configuration trained with contextual information, but is still able to provide meaningful classification. Looking at the low-dimensional representations of model outputs shows that contextual information allows the model to make better distinctions between objects.

The classification problem presented in this work is a supervised learning problem, where the expected output of the classifier is known. With future surveys expected to discover orders of magnitudes more objects, a point of interest is identifying previously undiscovered objects - the `unknown unknowns'. Unsupervised learning is an approach where a model is trained to identify patterns within unlabelled data with minimal human intervention. This approach can be used to identify outliers within the data, and this process is usually referred to as novelty or anomaly detection.

Unsupervised learning algorithms have been used to identify anomalies within the Open Supernova Catalog \citep{2019pruzhinskaya}, and in conjunction with deep neural network supervised learning for anomaly detection in variable stars \citep{2019tsang}. Active learning has also been used for anomaly detection, and to identify informative objects for labelling to improve a learning model as new observations become available \citep{2019ishidaAAD}.

Within the GOTO discovery pipeline, an anomaly detection algorithm can be used once an object has been identified as real to detect potential new discoveries, or applied to subsets of classified objects to identify peculiarities within known classes. The high-dimensional output of the dense layer carries some feature representation of the data, and could be fed into an anomaly detection algorithm. Both supervised and unsupervised classifications are useful in astronomical surveys: supervised classification provides utility and automation by classifying new objects into known classes, and unsupervised classification acts as a facilitator for specific science goals that utilise observations of rare and novel objects or even discovery of new transients.

\section*{Acknowledgements}

%The Acknowledgements section is not numbered. Here you can thank helpful
%colleagues, acknowledge funding agencies, telescopes and facilities used etc.

We thank the anonymous referee for their comments in helping to improve the quality of this paper. The research of U.F.B and J.R.M are funded through a Royal Society PhD studentship (Royal Society Enhancement Award RGF\textbackslash EA\textbackslash180234) and University Research Fellowship, (Royal Society URF UF150689 and STFC grant ST/R000964/1) respectively. V.S.D and M.J.D acknowledge the support of a Leverhulme Trust Research Project Grant. R.P.B, M.R.K, and D.M.S acknowledge support from the ERC under the European Union’s Horizon 2020 research and innovation programme (grant agreement no. 715051; Spiders). R.L.C.S acknowledges funding from STFC Parts of this research were conducted by the Australian Research Council Centre of Excellence for Gravitational Wave Discovery (OzGrav), through project number CE170100004.

The Gravitational-wave Optical Transient Observer (GOTO) project acknowledges the support of the Monash-Warwick Alliance; Warwick University; Monash University; Sheffield University; the University of Leicester; Armagh Observatory \& Planetarium; the National Astronomical Research Institute of Thailand (NARIT); the University of Turku; the University of Manchester;  the University of Portsmouth; the Instituto de Astrof\'{i}sica de Canarias (IAC) and the Science and Technology Facilities Council (STFC). 

U.F.B would like to thank Maurico A. \`Alavarez and Fariba Yousefi for helpful discussions in producing the work presented in this paper.  

%%%%%%%%%%%%%%%%%%%%%%%%%%%%%%%%%%%%%%%%%%%%%%%%%%
\section*{Data Availability}

Data products will be available as part of planned GOTO public data releases.



\bibliographystyle{mnras}
\bibliography{bibliography} % if your bibtex file is called example.bib

%\begin{thebibliography}{99}
%\end{thebibliography}

%%%%%%%%%%%%%%%%%%%%%%%%%%%%%%%%%%%%%%%%%%%%%%%%%%

%%%%%%%%%%%%%%%%% APPENDICES %%%%%%%%%%%%%%%%%%%%%
\appendix

\section{Training and validation loss graphs}

\begin{figure*}
    \centering
    \begin{subfigure}[b]{0.34\textwidth}
         \centering
         \includegraphics[width=\textwidth]{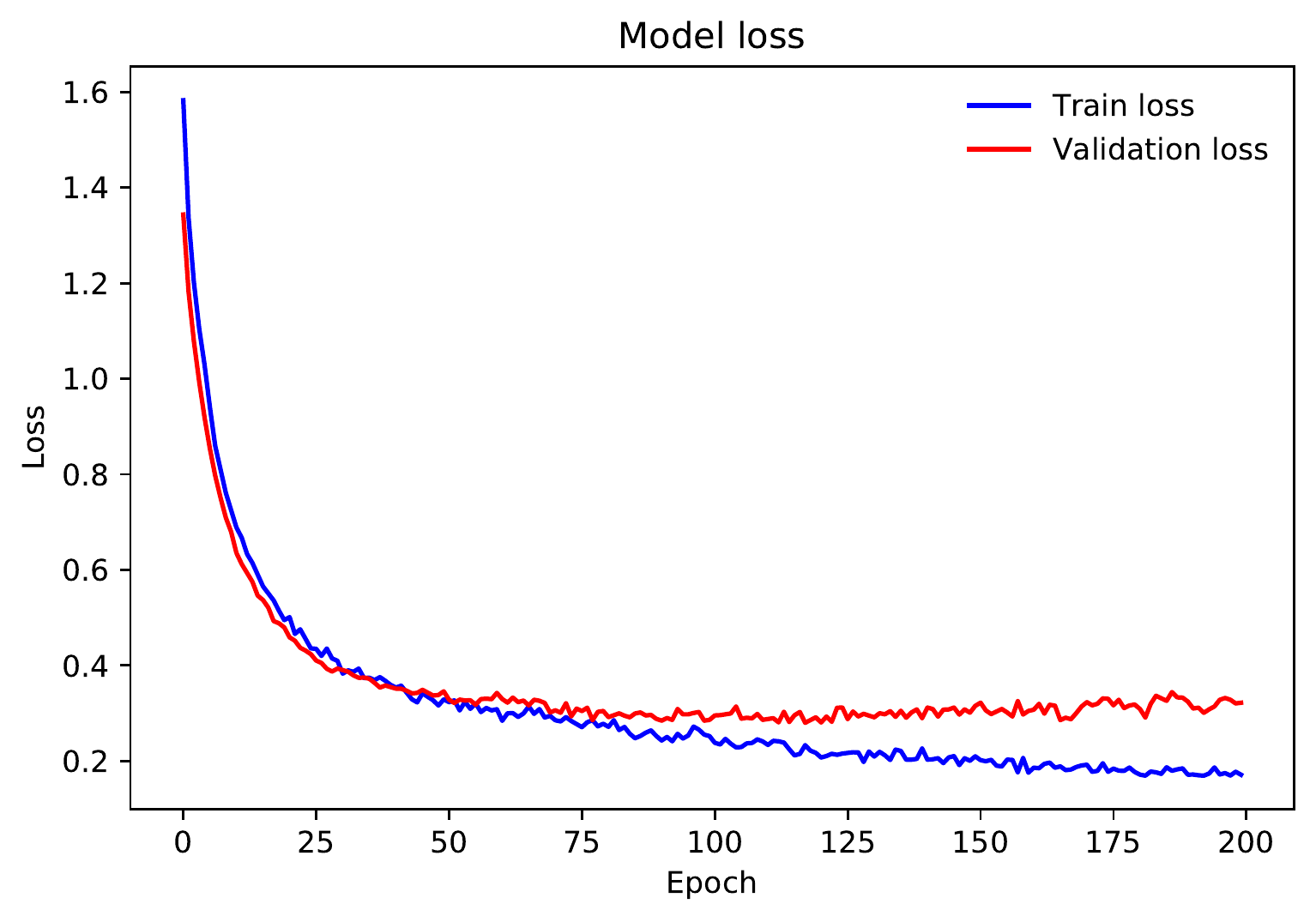}
         \caption{GRU with weighted focal loss.}
         \label{fig:lstm_fl_loss}
     \end{subfigure}
     \hspace{1.cm}
     \begin{subfigure}[b]{0.34\textwidth}
         \centering
         \includegraphics[width=\textwidth]{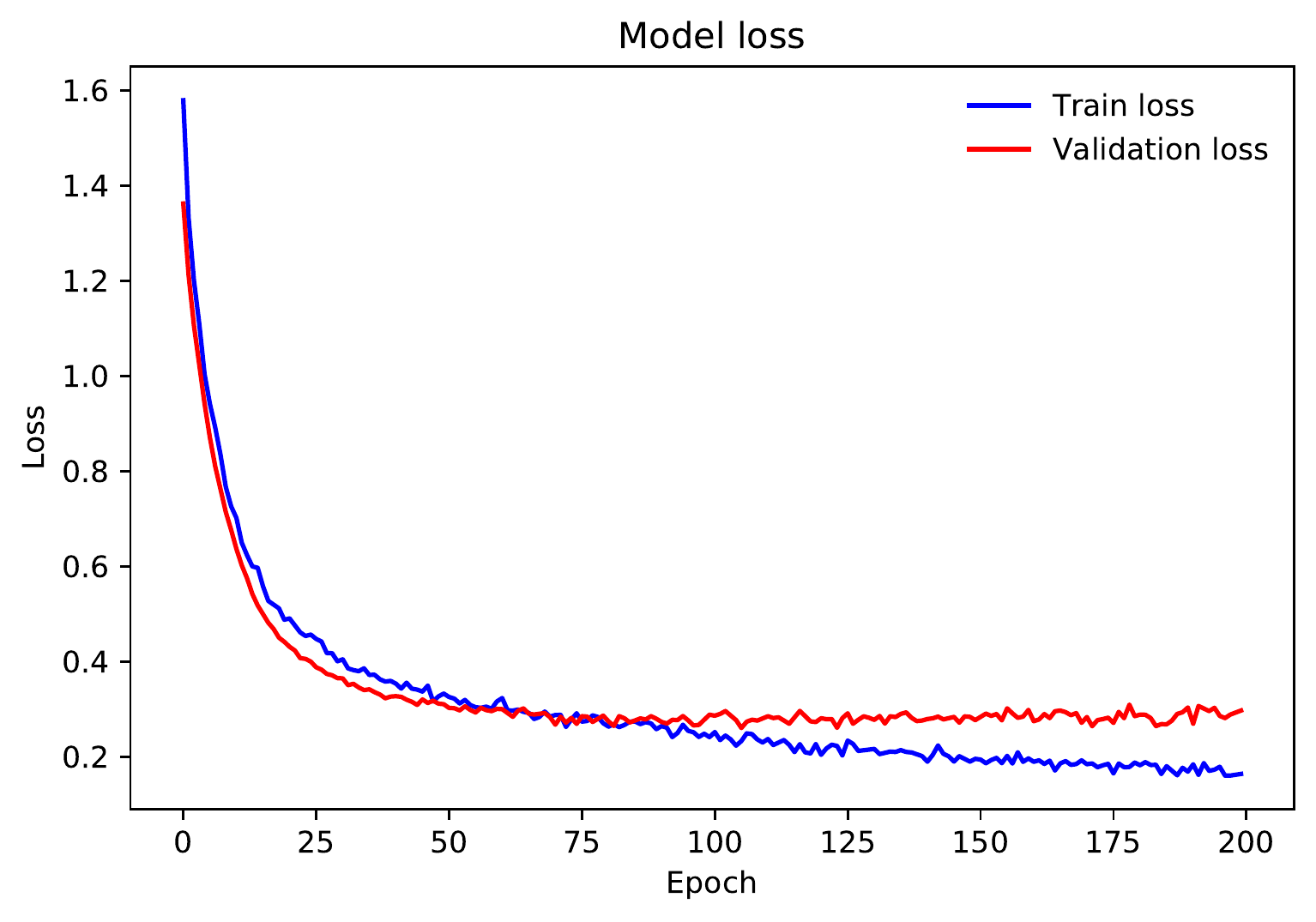}
         \caption{LSTM with weighted focal loss.}
         \label{fig:gru_fl_loss}
    \end{subfigure}
    
    \begin{subfigure}[b]{0.34\textwidth}
         \centering
         \includegraphics[width=\textwidth]{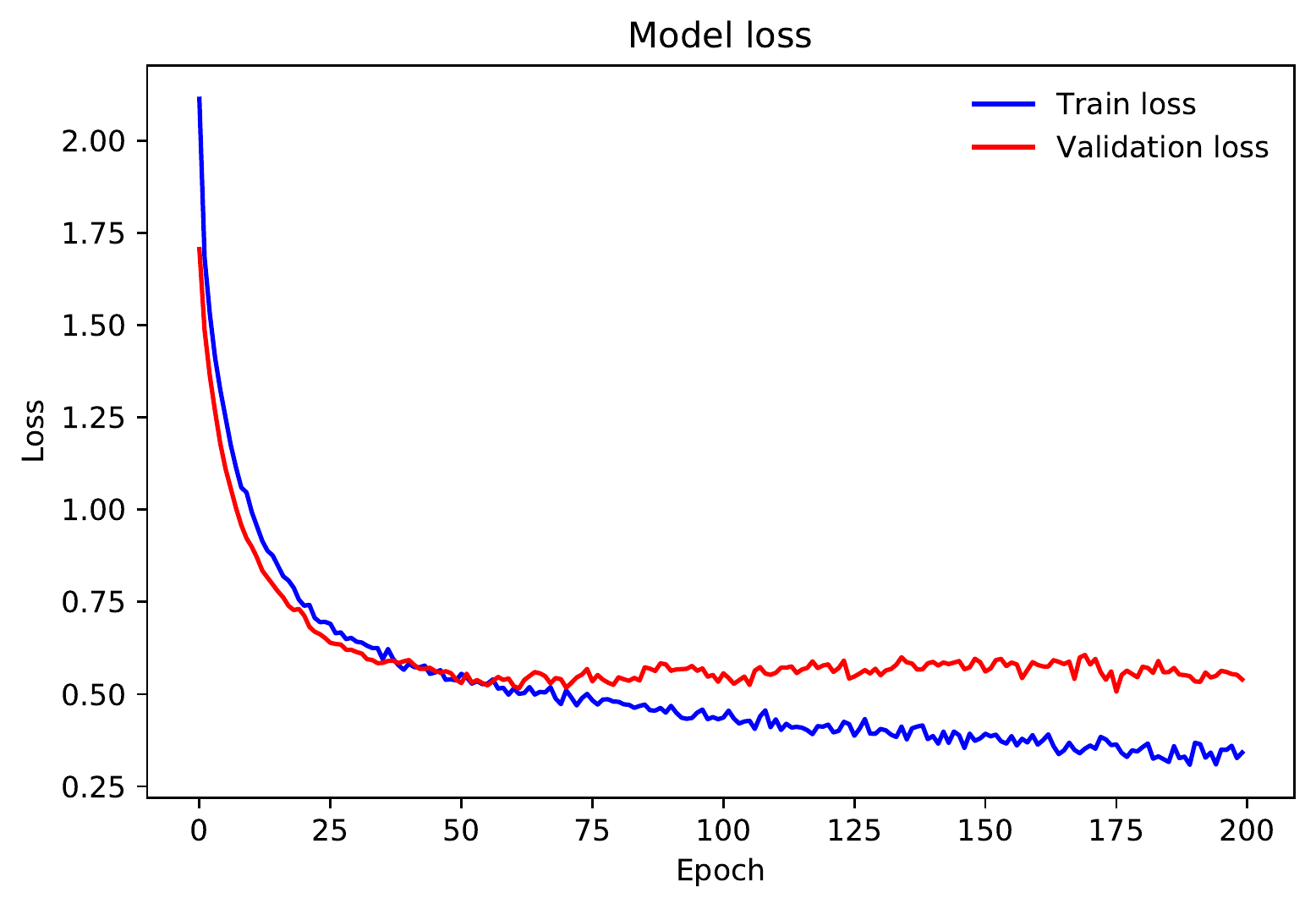}
         \caption{GRU with weighted cross entropy loss.}
         \label{fig:gru_wce_loss}
     \end{subfigure}
     \hspace{1.cm}
     \begin{subfigure}[b]{0.34\textwidth}
         \centering
         \includegraphics[width=\textwidth]{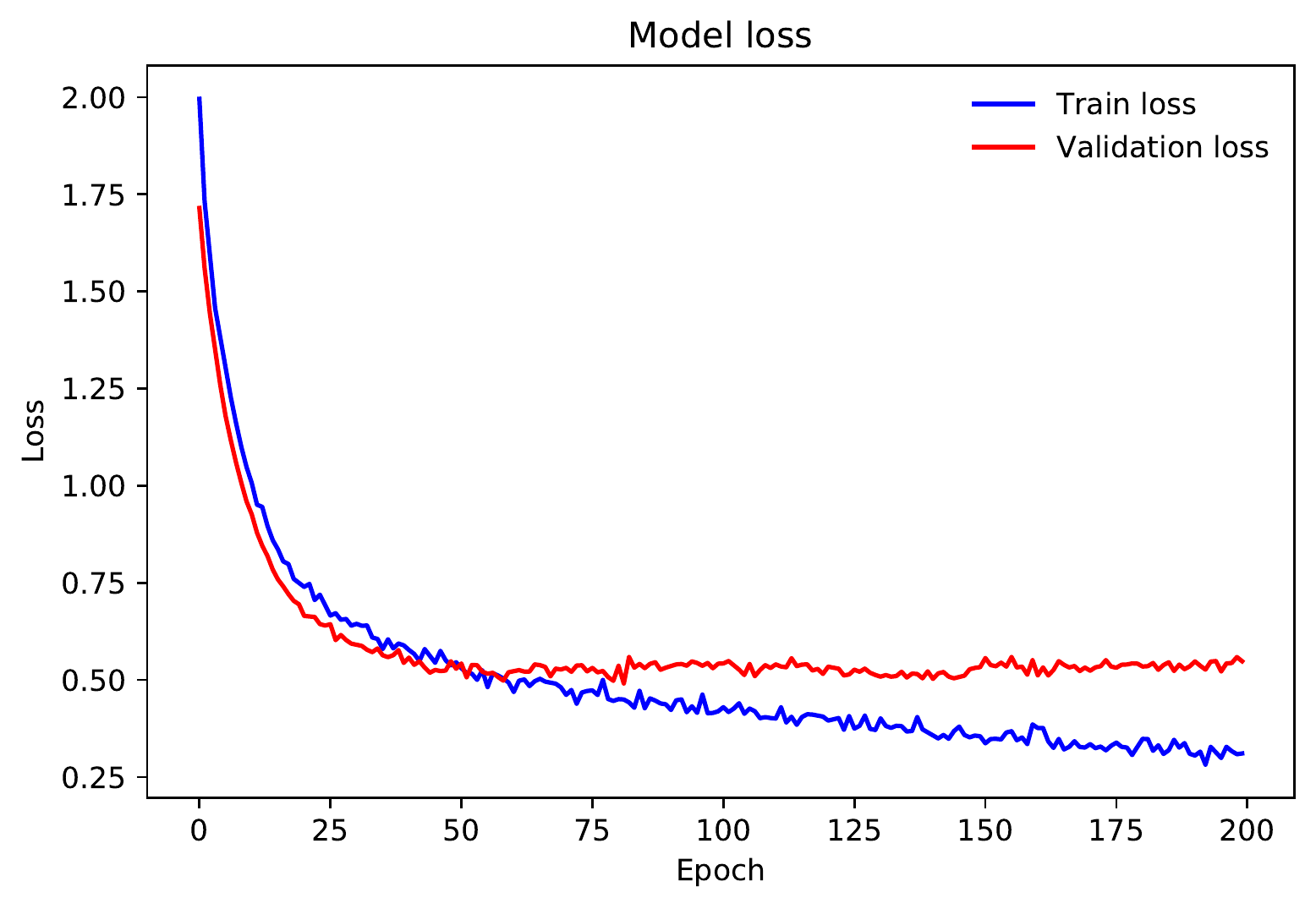}
         \caption{LSTM with weighted cross entropy loss.}
         \label{fig:lstm_wce_loss}
    \end{subfigure}
    
    \begin{subfigure}[b]{0.34\textwidth}
         \centering
         \includegraphics[width=\textwidth]{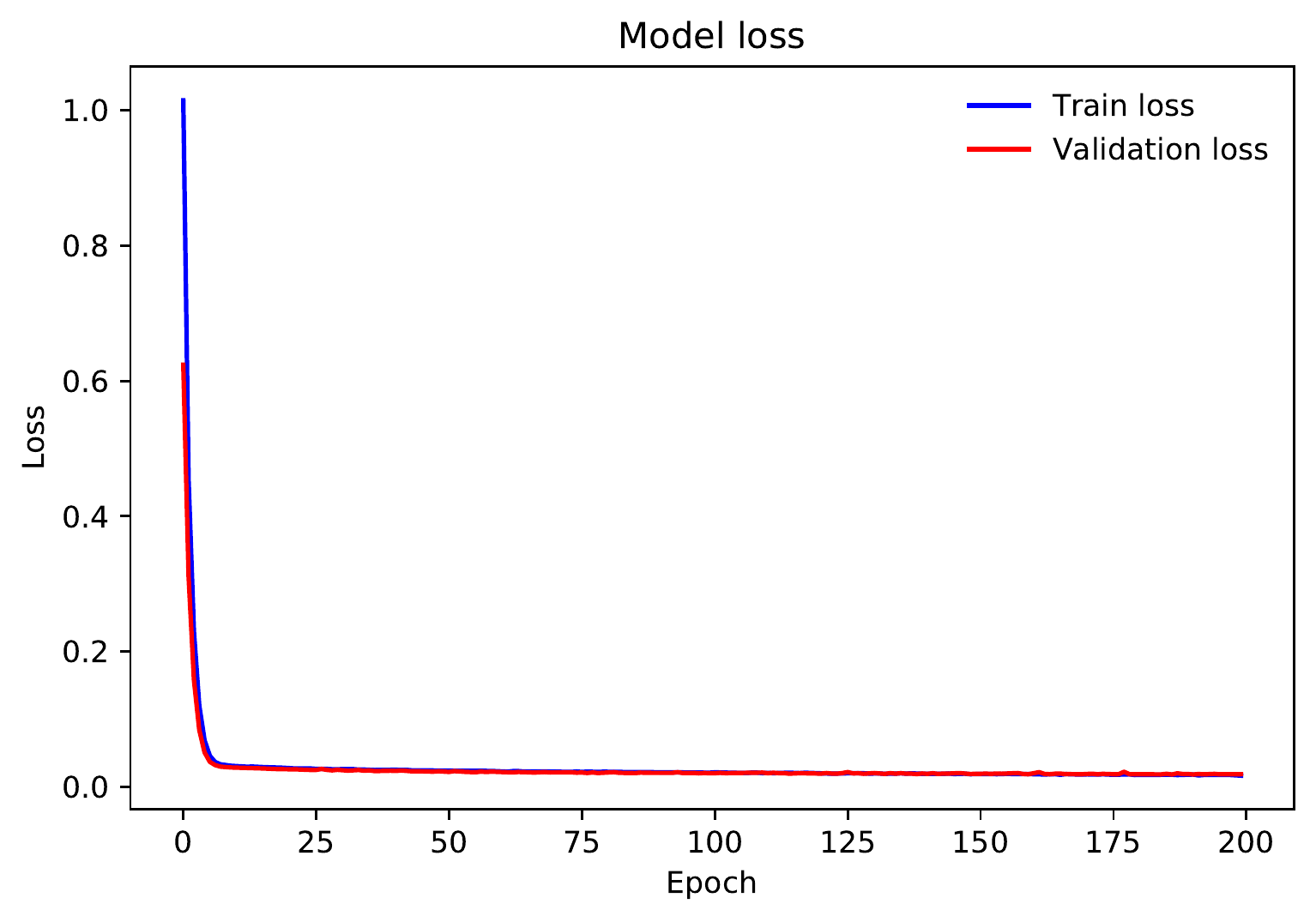}
         \caption{GRU with unweighted cross entropy loss.}
         \label{fig:gru_ce_loss}
     \end{subfigure}
     \hspace{1.cm}
     \begin{subfigure}[b]{0.34\textwidth}
         \centering
         \includegraphics[width=\textwidth]{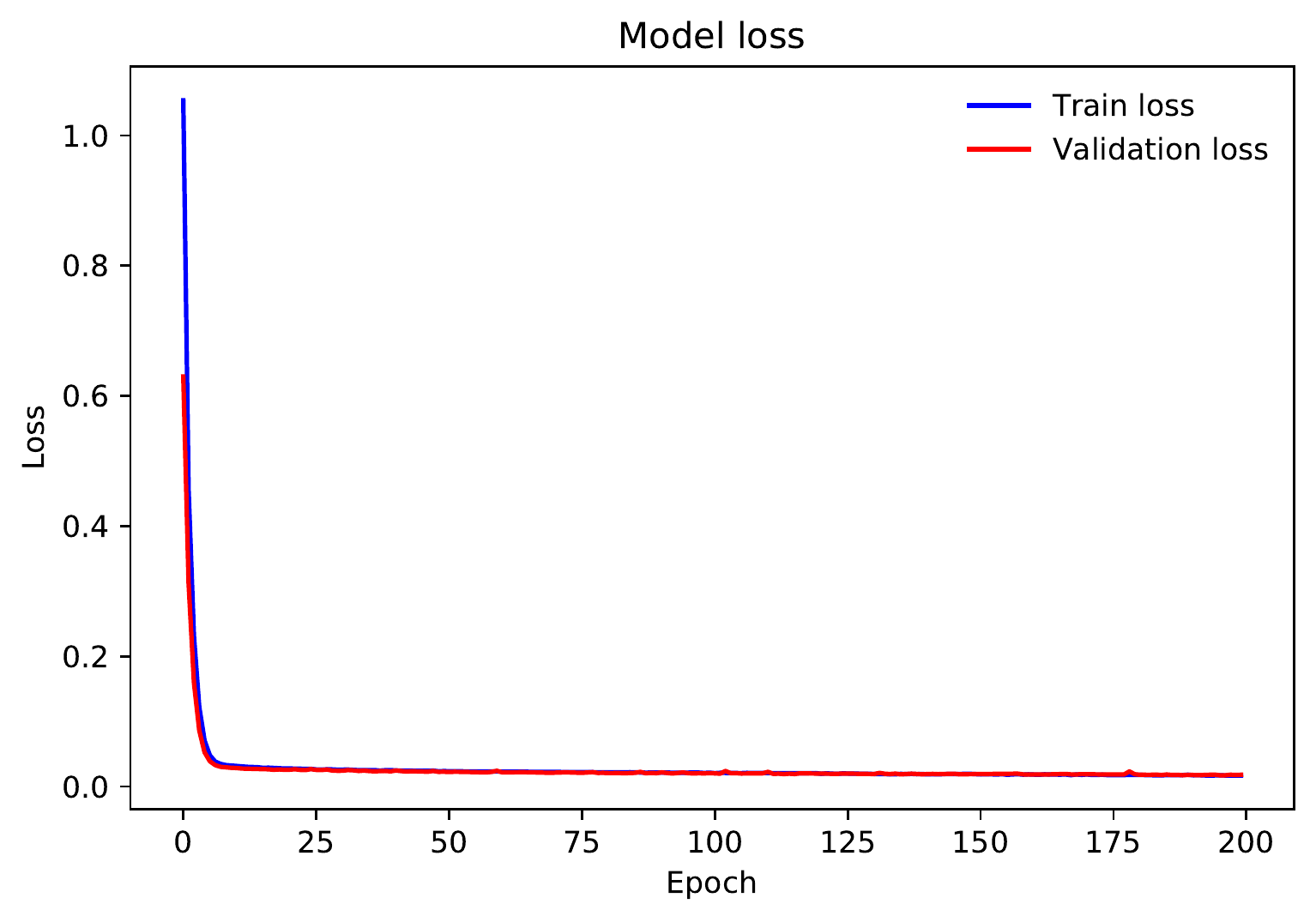}
         \caption{LSTM with unweighted cross entropy loss.}
         \label{fig:lstm_ce_loss}
    \end{subfigure}
    
    \centering
    \begin{subfigure}[b]{0.34\textwidth}
        \includegraphics[width=\textwidth]{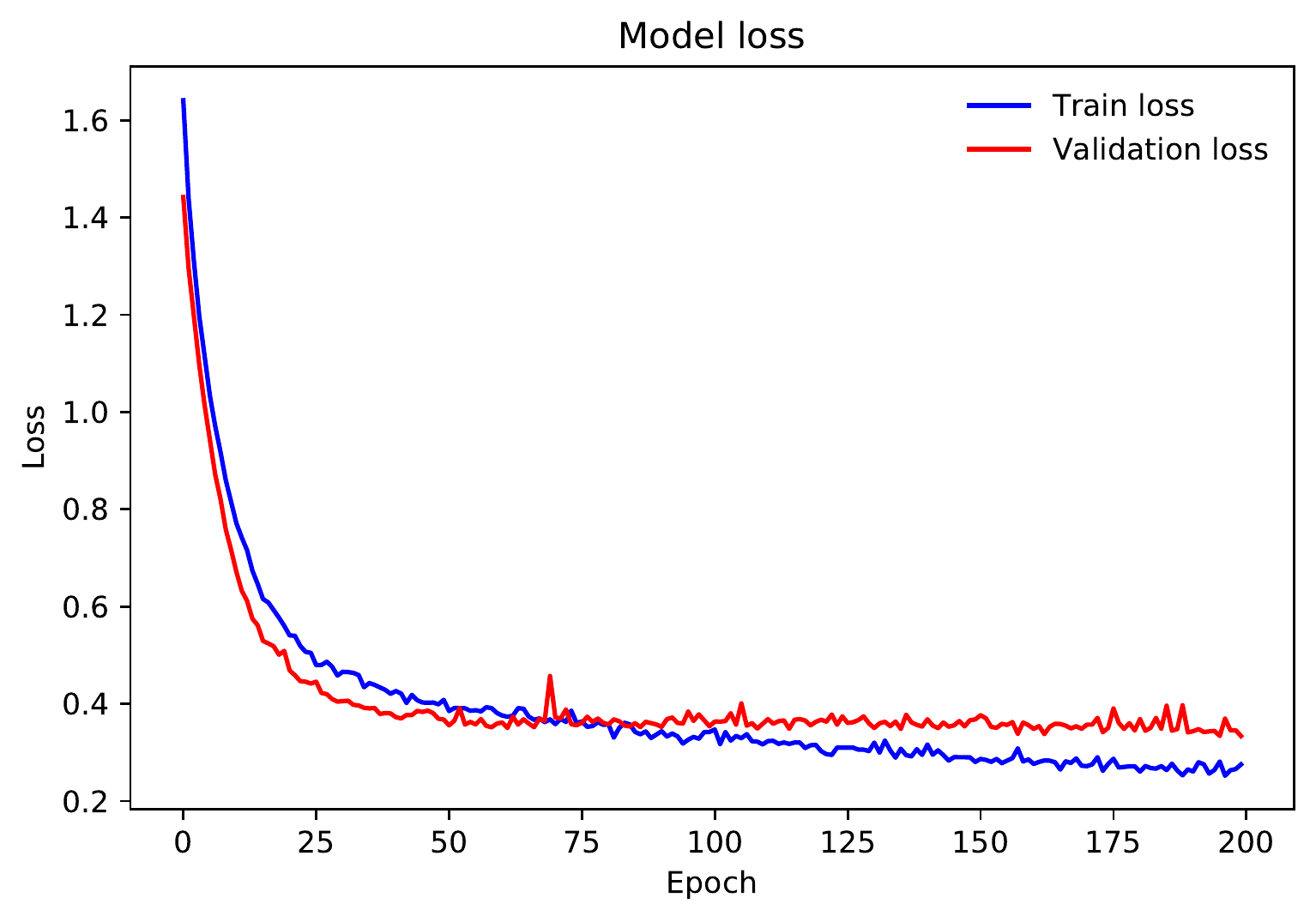}
        \caption{GRU with weighted focal loss, trained only time-series data}
        \label{fig:gru_fl_timeonly_loss}
    \end{subfigure}
    
    \caption{Evolution of training and validation loss for the best performing models during training. Models with the cross entropy loss converge quickly, but the loss is dominated by contribution from easy to classify examples. Models with the weighted cross entropy loss and focal loss eventually converge within 200 epochs of training, and are also able to account for examples from the minority classes.}
    \label{fig:train_loss}
\end{figure*}

Figure \ref{fig:train_loss} show how the models optimize the different loss functions during training. At each epoch, the model weights are updated through gradient descent such that the loss will be minimized. A model is said to converge once the the loss stops decreasing, indicating that the model has reached a local minimum in weight space.

\section{t-Distributed Stochastic Neighbor Embedding}

t-SNE is a dimensionality reduction technique, that takes a high-dimensional data set $\mathcal{X}=\{x_1, x_2, ..., x_n\}$ and converts it into a low-dimensional representation  $\mathcal{Y}=\{y_1, y_2, ..., y_n\}$. The low-dimensional data points $\mathcal{Y}$ are mappings of the high-dimensional data points $\mathcal{X}$ in the low-dimensional space.

The high-dimensional Euclidean distances between points are converted into probabilities that represent similarities between points by centering a Gaussian distribution onto each point. The similarity between points $x_i$ and $x_j$ are encapsulated in the joint probability $p_{ij}$; if $x_i$ and $x_j$ are near, then $p_{ij}$ will be high and if they are far apart then $p_{ij}$ will be low.  

In the low-dimensional mapping, the probability $q_{ij}$ is a measure of similarity between $y_i$ and $y_j$, and $q_{ij}$ will be high if the two points are near each other and low if they are far apart. Instead of centering a Gaussian distribution onto the points in low-dimensional space, a student t-distribution is used instead. Using a heavier-tailed distribution allows moderate distances in the high-dimensional map to be modelled by larger distances in the lower-dimensional map, preventing a 'crowding' of points that are not too dissimilar \citep{2008vandermaaten}.

The variance of the Gaussian distribution is set so that the probability distribution produced by the variance has a fixed perplexity, which is set by the user. The perplexity can be thought of as a measure of the effective number of neighbors in the region of the data point in question. Here, the perplexity parameter is set to 20.

Given the two joint probabilities $p_{ij}$ and $q_{ij}$, t-SNE determines an optimal low-dimensional mapping $\mathcal{Y}$ of the high-dimensional data set $\mathcal{X}$ by minimising the Kullback-Leibler divergence \citep{kullback1951} of $q_{ij}$ and $p_{ij}$ using a gradient descent method.

%%%%%%%%%%%%%%%%%%%%%%%%%%%%%%%%%%%%%%%%%%%%%%%%%%

% Don't change these lines
\bsp	% typesetting comment
\label{lastpage}
\end{document}